# Highly Tunable Magnetic and Magnetotransport Properties of Exchange Coupled Ferromagnet/Antiferromagnet-based Heterostructures

Sri Sai Phani Kanth Arekapudi,*[a] Daniel Bülz,[a] Fabian Ganss,[a] Fabian Samad,[a, b] Chen Luo,[c,d] Dietrich R. T. Zahn,[a] Kilian Lenz,[b] Georgeta Salvan,[c] Manfred Albrecht,[e] and Olav Hellwig[a, b]

[a]Institute of Physics, Technische Universität Chemnitz, 09107 Chemnitz, Germany.

[b]Institute of Ion Beam Physics and Materials Research, Helmholtz-Zentrum Dresden-Rossendorf, Bautzner Landstrasse 400, 01328 Dresden, Germany.

[c]Helmholtz-Zentrum Berlin für Materialien und Energie, Albert-Einstein-Strasse 15, 12489 Berlin, Germany.

[d]Institute of Experimental Physics of Functional Spin Systems, Technical University Munich, James-Franck-Str. 1, 85748, Garching b. München, Germany.

[e]Institute of Physics, University of Augsburg, Universitätsstraße 1, 86159 Augsburg, Germany.

**ABSTRACT:** Antiferromagnets (AFMs) with zero net magnetization are proposed as active elements in future spintronic devices. Depending on the critical thickness of the AFM thin films and the measurement temperature, bimetallic Mn-based alloys and transition metal oxide-based AFMs can host various coexisting ordered, disordered, and frustrated AFM phases. Such coexisting phases in the exchange coupled ferromagnetic (FM)/AFM-based heterostructures can result in unusual magnetic and magnetotransport phenomena. Here, we integrate chemically disordered AFM $\gamma$-IrMn$_3$ thin films with coexisting AFM phases into complex exchange coupled MgO(001)/$\gamma$-Ni$_3$Fe/$\gamma$-IrMn$_3$/$\gamma$-Ni$_3$Fe/CoO heterostructures and study the structural, magnetic, and magnetotransport properties in various magnetic field cooling states. In particular, we unveil the impact of rotating the relative orientation of the disordered and reversible AFM moments with respect to the irreversible AFM moments on the magnetic and magnetoresistance properties of the exchange coupled heterostructures. We further found that the persistence of AFM grains with thermally disordered and reversible AFM order is crucial for achieving highly tunable magnetic properties and multi-level magnetoresistance states. We anticipate that the introduced approach and the heterostructure architecture can be utilized in future spintronic devices to manipulate the thermally disordered and reversible AFM order at the nanoscale.

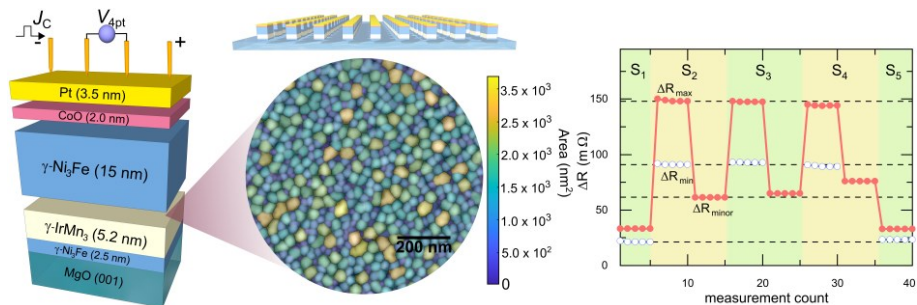

**Abstract Figure**

# 1. INTRODUCTION

Spintronic devices based on antiferromagnetic materials for data storage and signal processing offer robustness against large magnetic fields,[1-3] exchange energy enhanced terahertz writing speeds,[4] and no net magnetization ($M = 0$); hence, surpass dipolar-field limited scalability.[5] In an AFM with two equivalent sublattice magnetizations denoted by $\mathbf{M}_A$ and $\mathbf{M}_B$ oriented anti-parallel, $|\mathbf{M}_A| = |\mathbf{M}_B| = M_0$ is the magnitude of the sublattice magnetization; the normalized AFM Néel vector (order parameter) is given by $\mathbf{l} = (\mathbf{M}_A - \mathbf{M}_B)/(2M_0)$.[6] Electrical manipulation of the AFM Néel vector with the prospect of encoding information was achieved in various AFM compounds, such as CuMnAs,[2] Mn$_2$Au,[7] Mn$_3$Pt,[5,8] Mn$_3$Sn,[9,10] IrMn$_3$,[11] NiO,[12] and CoO,[13] via a relativistic mechanism, namely Néel order spin-orbit torque[14] (SOT)— presumably without invoking heat or magnetic fields. In strong contrast, recent experiments showed that significant amounts of charge current densities ($\geq 4 \times 10^7$ A/cm$^2$) beyond the linear ohmic regime of the devices are needed to switch the Néel vector (or the local AFM domains).[15-18] The power efficiency, the physical mechanism of switching, and the efficiency of the electrical detection method for validating the current-induced microscopic switching of the AFM domains are currently under investigation.[17-20] Moreover, recent work based on NiO/Pt bilayer devices revealed a current-induced purely thermomagnetoelastic switching mechanism.[20]

Alternatively, the AFM Néel vector orientation can be manipulated by magnetic field cooling (MFC) the AFM or FM/AFM based heterostructures below the Néel temperature $T_N$ of the AFM.[21-24] In particular, field cooling the FM/AFM heterostructures below $T_N$ and blocking temperature $T_B$ results in a shift of the FM (Ni$_3$Fe) hysteresis loop along the magnetic field axis known as the exchange bias (EB) field, which arises from the interfacial exchange coupling between the AFM and FM layers.[24,25] Earlier, theoretical models and experimental studies have extensively investigated the magnetic, magnetotransport and switching properties of the prototypical exchange coupled FM/AFM heterostructures.[1,22,24-30] The EB effect in FM/AFM hetero-structures is attributed to a minuscule amount of irreversible and uncompensated interfacial AFM spins created by the inevitable site disorder at an atomic scale.[31,32] In contrast, the reversible and uncompensated interfacial AFM spins can cause an enhancement in the coercive field, EB training effect, and asymmetric magnetization reversal in the adjacent FM layer.[27,28,33-36] Recent experimental work has revealed a linear dependence of the EB field strength on the amount of pinned uncompensated interfacial spins in various exchange coupled FM/AFM heterostructures.[37] Due to their microscopic and elusive nature, manipulating the uncompensated interfacial AFM moments is typically considered challenging.[22] Additionally, coexisting ordered, disordered (paramagnetic), and frustrated phases in AFM materials can cause various unusual phenomena.[31,35,38-40] Therefore, an understanding and manipulation of the thermally disordered, reversible, and frustrated AFM order at the nanoscale in the exchange coupled systems is of central importance.[41-45] However, the role of coexisting AFM phases and fractional variation in the AFM order on the magnetic and magnetotransport properties of the FM/bimetallic AFM/FM/Transi-tion Metal (TM)-oxide heterostructures remains an outstanding challenge.

Chemically disordered $\gamma$-IrMn$_3$ thin films with collinear AFM spin structure are extensively employed in research and application; due to their ease of integration, giant charge-to-spin current conversion efficiency (up to 0.8),[46] high spin Hall conductivity (~$4\times10^3\hbar/e$ S/cm),[46] large magnetocrystalline anisotropy energy (-7.05 meV per unit cell),[47] and high $T_N$ of about 730 K.[48,49] Moreover, bimetallic IrMn alloys exhibit a strong correlation between chemical, structural, magnetic, and electronic properties–originating from the combination of the spontaneous magnetic moment of the Mn $3d$ shell and spin-orbit coupling (SOC) with the Ir $5d$ shell.[49-51]

A critical IrMn grain volume ($V > V_C$) is necessary at a given measurement temperature to establish a stable AFM order.[26,52] However, in granular AFM IrMn thin films, a rather broad (nanoscale) grain size distribution is typically observed.[26,52,53] Coexisting AFM order can cause highly undesirable properties in FM/AFM heterostructure based spintronic devices, such as the decay of interfacial SOT,[46,54] enhancement of the effective damping (or magnon relaxation rate) in the adjacent FM,[54] distorted (FM/AFM) interfacial spin structure,[31,55] interfacial spin memory loss,[56] non-linear operation, thermal dependence of the AFM anisotropy energy barrier,[26,52] EB training effects,[27,34,36,57] coercive field enhancement,[25,28,35,38] as well as unusual magnetotransport phenomena.[1,23,30,43,46,54]

In this study, we integrate $\gamma$-phase IrMn$_3$ thin films with coexisting AFM phases into epitaxially textured MgO(001)/ Ni$_3$Fe(2.5 nm)/IrMn$_3$(5.2 nm)/Pt(3.5 nm) and MgO(001)/ Ni$_3$Fe(2.5 nm)/IrMn$_3$(5.2 and 16.0 nm)/Ni$_3$Fe(15 nm)/ CoO(0 and 2.0 nm)/Pt(3.5 nm) heterostructures and investigate the structural, magnetic, and magnetotransport properties in various MFC states. We identify two distinct MFC-dependent magnetization reversal and magnetotransport regimes depending on the IrMn$_3$ film thickness $d_{IrMn3}$ in the exchange coupled heterostructures. In the first regime, with $d_{IrMn3}$ = 5.2 nm, we reveal highly responsive (or active) MFC-dependent magnetization reversal and magnetotransport properties. We manipulate the relative orientation of the thermally disordered and reversible AFM moments with respect to the irreversible AFM moments via an iterative MFC procedure and achieve highly tunable magnetization reversal properties and multi-level magnetoresistance. In the latter regime, with $d_{IrMn3}$ = 16.0 nm ($V \gg V_C$), we reveal MFC-independent (or rigid) magnetization reversal properties with stable EB field and fewer magnetoresistance states. However, we show that inserting an ultra-thin AFM CoO(2.0 nm) layer with a fully disordered/reversible AFM order enables a certain degree of tunability in the EB field strength and MFC responsive magnetization reversal behavior. Moreover, MgO(001)/FM/bimetallic AFM/FM/TM-oxide heterostructures can be readily integrated into future spintronic devices to realize multi-level resistance states,[43] active AFM-based tunnel junction devices,[1,23] and magnon gates.[58,59]

Details regarding the sample preparation and experimental techniques can be found in the experimental section. Supplementary synchrotron-based element-specific X-ray absorption (XAS) and X-ray magnetic circular dichroism (XMCD) measurements on similar, Si$_3$N$_4$/Ta(3.5 nm)/Ni$_3$Fe (1.5 nm)/IrMn$_3$(16.0 nm)/Ni$_3$Fe(10.0 nm)/Co(5.0 nm)/ CoO(2.0 nm)/TaOx(2.5 nm) heterostructures provide evidence for the modified (interfacial and bulk) AFM spin configuration in various MFC states. Additionally, magnetic (XMCD) hysteresis loops obtained at the Ni, Co, and Mn $L_3$-edges offer direct insights into the magnetization reversal behavior and the underlying interfacial properties of the exchange coupled Ni/Co, Mn/Ni, and Co/CoO interfaces in various MFC states.

## 2. RESULTS AND DISCUSSION

**2.1. Iterative Magnetic Field Cooling Procedure.** We consider a simple exchange coupled MgO(001)/Ni$_3$Fe(2.5 nm)/ IrMn$_3$(5.2 nm) heterostructure to introduce the iterative MFC procedure and demonstrate the tunable magnetization reversal and magnetotransport properties. Due to the substantial in-plane lattice mismatch (~16%) between the single crystal MgO(001) substrate and Ni$_3$Fe(2.5 nm) thin films, we expect a large grain volume distribution in Ni$_3$Fe(2.5 nm)/IrMn$_3$(5.2 nm) thin films. As shown in Figure 1(a) (steps 1 and 2), we field annealed the sample at $T_A = 605$ K for 90 min in +7 T magnetic field (field annealing process), followed by the MFC down to the configuration temperature $T_{config} = 300$ K. The MFC procedure results in a shift of the FM (Ni$_3$Fe) hysteresis loop along the magnetic field axis known as the EB field $H_{eb}$.[24-26] The AFM IrMn$_3$ grains that are ordered and thermally stable at a given measurement temperature can contribute to $H_{eb}$.[26,52,53] The IrMn$_3$ grains with $V<V_C$ can be fully disordered and may act as non-magnetic defects within the AFM thin film.[26,52,53] Due to the inevitable structural disorder, a fraction of the IrMn$_3$ grains with volume in the vicinity of $V_C$ that are exchange decoupled from the neighboring IrMn$_3$ grains can possess a reversible AFM order.[26,53,60]

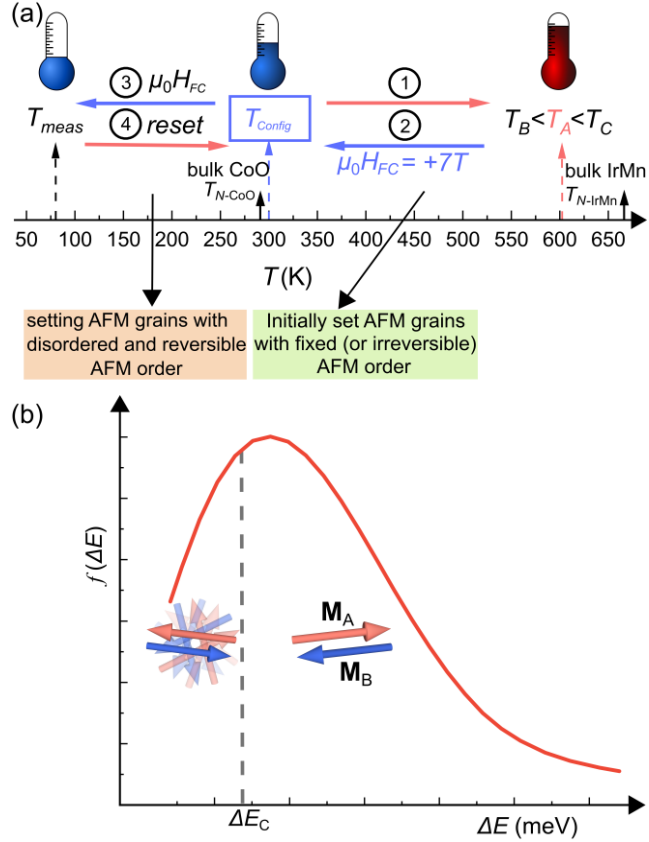

**Figure 1. (a)** The four-step (iterative) MFC procedure with the configuration temperature $T_{config}$ tailored to the AFM IrMn and CoO thin films investigated in this study. **(b)** Schematic of the thermally activated energy barrier distribution of the granular AFM thin films, where $\Delta E_C$ is the critical anisotropy energy barrier.[26,52,60] The iterative MFC process in (a) steps 3 and 4 can be used to set the relative orientation of the AFM moments with $\Delta E<\Delta E_C$ (either parallel or anti-parallel) with respect to the AFM grains ($\Delta E>\Delta E_C$) with irreversible AFM order.

In exchange coupled FM/AFM heterostructures, $T_N$ of the AFM and $T_B$ show an AFM grain volume-dependent distribution (or scaling).[26,52,60] By cooling the exchange coupled heterostructures $T<T_{Config}$ in a magnetic field (step 3 in Figure 1(a)), IrMn$_3$ grains with $V<V_C$ achieve spontaneous AFM-order and contribute to $H_{eb}$. Upon transition from a disordered to an ordered AFM phase, the combination of atomic site disorder and local AFM exchange interactions between the adjacent Mn atoms can result in various energetically degenerate spin configurations or frustrated AFM order.[31,38-40] The presence of such frustrated AFM order in FM/AFM heterostructures can cause an enhancement in the coercive field $H_c$ and an undesirable EB training effect.[35]

The AFM anisotropy energy barrier is given by $\Delta E = K_{AFM}V$, where $K_{AFM}$ is the anisotropy energy density of the AFM thin film.[26] The anisotropy energy barrier distribution $f(\Delta E)$ is dependent on the grain volume distribution and closely follows a log-normal distribution,[60] as shown in Figure 1(b). The AFM grains with $\Delta E > \Delta E_C$ exhibit a fixed (or irreversible) Néel order after the field annealing process (steps 1 and 2 in Figure 1(a)). In contrast, AFM grains with $\Delta E < \Delta E_C$ possess a disordered/reversible AFM order when $T > T_{Config}$.

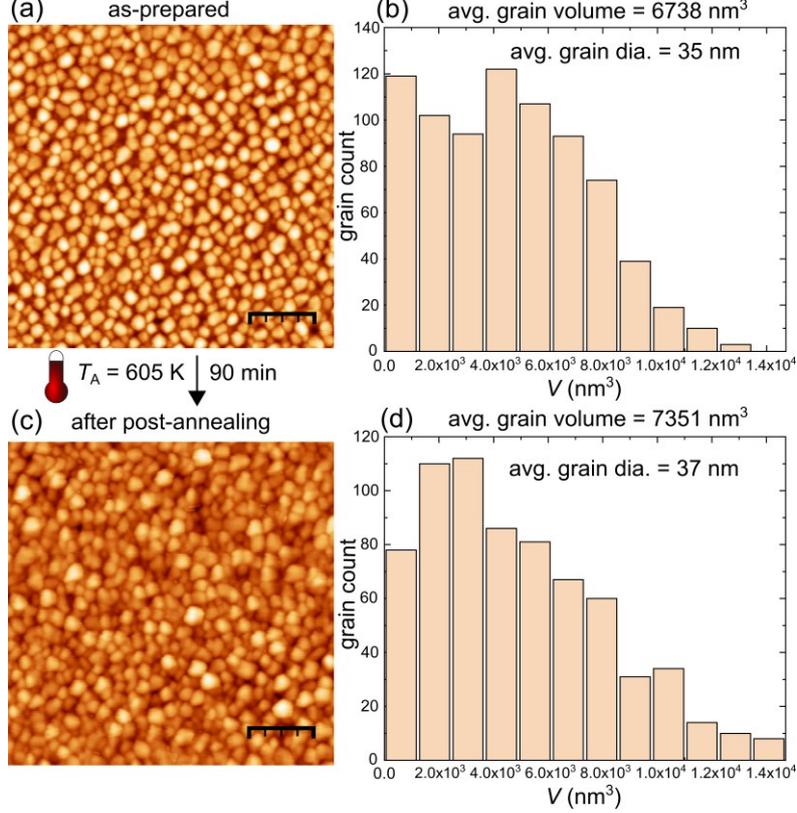

**Figure 2.** Grain volume distribution of epitaxially textured MgO(001)/Ni$_3$Fe(2.5 nm)/IrMn$_3$(5.2 nm) sample imaged at room temperature. **(a)** Atomic force microscopy scan of the sample surface, and **(b)** the corresponding grain volume distribution in the as-deposited state. **(c)** The sample surface after the field annealing process at 605 K for 90 min in +7 T magnetic field and subsequent field cooling down to 300 K, and **(d)** the corresponding grain volume distribution. Scale bar in (a) and (c) is 200 nm.

In a simple exchange coupled FM/AFM system with $T < T_{Config}$, the EB field in the case of opposite Néel vector alignment between the AFM grains with $V > V_C$ and $V < V_C$ is given by

$$\mu_0 H_{eb} \propto \int_{\Delta E_C}^{\infty} f(\Delta E)d(\Delta E) - \int_0^{\Delta E_C} f(\Delta E)d(\Delta E), \quad (1)$$

the first term denotes the contribution from the initially set (or irreversible) AFM grains with $V > V_C$, and the second term denotes the contribution from the AFM grains with $V < V_C$.[26,52,60] By appropriate selection of the sign and magnitude of the cooling field $H_{FC}$ in step 3 (see Figure 1(a)), the Néel vector orientation of the reversible and disordered AFM grains with $V \leq V_C$ can be aligned either parallel or anti-parallel with respect to the initially set (irreversible) AFM grains ($V > V_C$); thereby tuning the sign and EB field strength at a given measurement temperature $T_{meas}$. A *reset* operation can be performed simply by heating the sample back to 300 K.

**2.2. Grain Volume Distribution of IrMn₃ thin film.** Figures 2(a) and 2(c) show the surface of the MgO(001)/Ni₃Fe (2.5 nm)/IrMn₃(5.2 nm) sample before and after the field annealing process as imaged by atomic force microscopy, and the corresponding IrMn₃ grain volume distribution shown in Figures 2(b) and 2(d), respectively. The volume distribution is calculated from the surface area distribution of the IrMn₃ grains, assuming a full structural coherence length of the individual grains. After the field annealing process, there is a decrease in the number of grains with $V<1.0 \times 10^3$ nm$^3$ (see Figure 2(d)). The IrMn₃ grains below a critical volume $V_C$ ($<<2.0 \times 10^3$ nm$^3$) are paramagnetic; they act as non-magnetic defects at room temperature.[53] These defects might hinder the intergranular exchange coupling and thus reduce the AFM domain size in IrMn₃ thin films.[53,61,62] We obtain an average grain diameter of ~35.0(±2.0) nm in an as-deposited state, estimated from the equivalent diameter of more than 750 grains.

Similarly, we found a marginal increase in the average grain diameter of ~37.0(±2.5) nm after the field annealing process. It is worth noting that the average grain diameter remains similar for the samples with IrMn₃ film thickness $d_{IrMn3}$ = 16 nm, indicating a columnar grain growth with almost no coalescence of grains with increasing film thickness. In the X-ray diffraction (XRD) experiments, the absence of the diffraction peaks corresponding to the $L1_2$ phase (superstructure) of IrMn₃ and Ni₃Fe thin films confirms a chemically disordered γ-phase (see Figure 5(f)). We obtain the structural coherence length $\tau_{IrMn} \approx 5.0(\pm0.3)$ nm along the out-of-plane orientation of the IrMn₃ layer using the Scherrer equation. Therefore, we conclude that most of the AFM IrMn₃ grains displayed in Figure 2(c) show full structural coherence length ($d_{IrMn} \cong \tau_{IrMn}$) along the growth direction.

**2.3. Magnetization Reversal and Magnetotransport Properties of MgO(001)/NiFe₃/IrMn₃ Heterostructures.** We now study the magnetic properties of the MgO(001)/ Ni₃Fe (2.5 nm)/IrMn₃ (5.2 nm)/Pt(3.5 nm) sample in various MFC states by applying a magnetic field along the [100] axis ($\mu_0 H$|[100]), as shown in Figure 3(a). Initially, we field annealed the samples at 605 K for 90 mins; subsequently, field cooled down to 300 K in +7 T magnetic field, as shown in Figure 1(a) steps 1 and 2. Figure 3(b) shows the simplified spin structure of the FM/AFM heterostructure at 300 K, including the AFM (IrMn₃) grains with irreversible (or fixed), reversible, and thermally disordered AFM phases.

We obtain a characteristic EB hysteresis loop that is shifted to the negative magnetic field direction after training the sample,[25,27] as shown in Figure 3(c). We extract $\mu_0 H_{eb} = 1/2(\mu_0 H_{c\text{-right}} + \mu_0 H_{c\text{-left}})$ = -20.05 mT and $\mu_0 H_C = 1/2(\mu_0 H_{c\text{-right}} - \mu_0 H_{c\text{-left}})$ = 18.90 mT. We obtain the effective interfacial exchange coupling constant $J_{eb} = -\mu_0 H_{eb} M_{FM} d_{FM}$ = 34.83(±2.8) μJ/m$^2$ at 300 K, where the FM Ni₃Fe(2.5 nm) layer magnetization $M_{FM}$ = 695 kA/m (obtained from SQUID-VSM measurements) and the layer thickness $d_{FM}$ = 2.5(±0.2) nm. When the exchange coupled heterostructure is uniformly field cooled in +7 T (from 605 K to 300 K), we expect the Néel vector orientation of most AFM IrMn₃ grains with $V>V_C$ to be in the same direction. We estimate the AFM anisotropy constant $K_{AFM}$, using the generalized Meiklejohn-Bean model proposed by Binek *et al.*[63,64] $K_{AFM} = J_{eb}/2\sqrt{2}d_{AFM}^{cr}$ = 3.08(±0.16) kJ/m$^3$ (or 0.308(±0.016) × 10$^5$ erg/cm$^3$). $d_{AFM}^{cr}$ = 4(±0.2) nm is the minimum thickness of the AFM layer at which $|\mu_0 H_{eb}|>0$ for a given measurement temperature ($T$ = 300 K). The anisotropy energy barrier at 300 K can be calculated from the average (IrMn₃) grain volume $V$ = 7.351 x 10$^{-24}$ m$^3$ (see Figure 2(c)), and with $K_{AFM}$ from above, we obtain $\Delta E = K_{AFM} V$ = 0.0226(±0.0011) aJ. We further determine the approximate value of the exchange stiffness for the AFM grains using the expression $A_{AFM} = 3k_B T_N/aZ$,[27] where $k_B$ is the Boltzmann constant, $Z$ is the number of nearest neighbors in the face-centered cubic (fcc) lattice, and $a$ is the lattice parameter. We obtain $A_{AFM} = 5.13 \times 10^{-12}$ J/m, considering $T_N = 550$ K, $Z = 12$, and $a = 3.70$ Å; consistent with the earlier reports.[27,65]

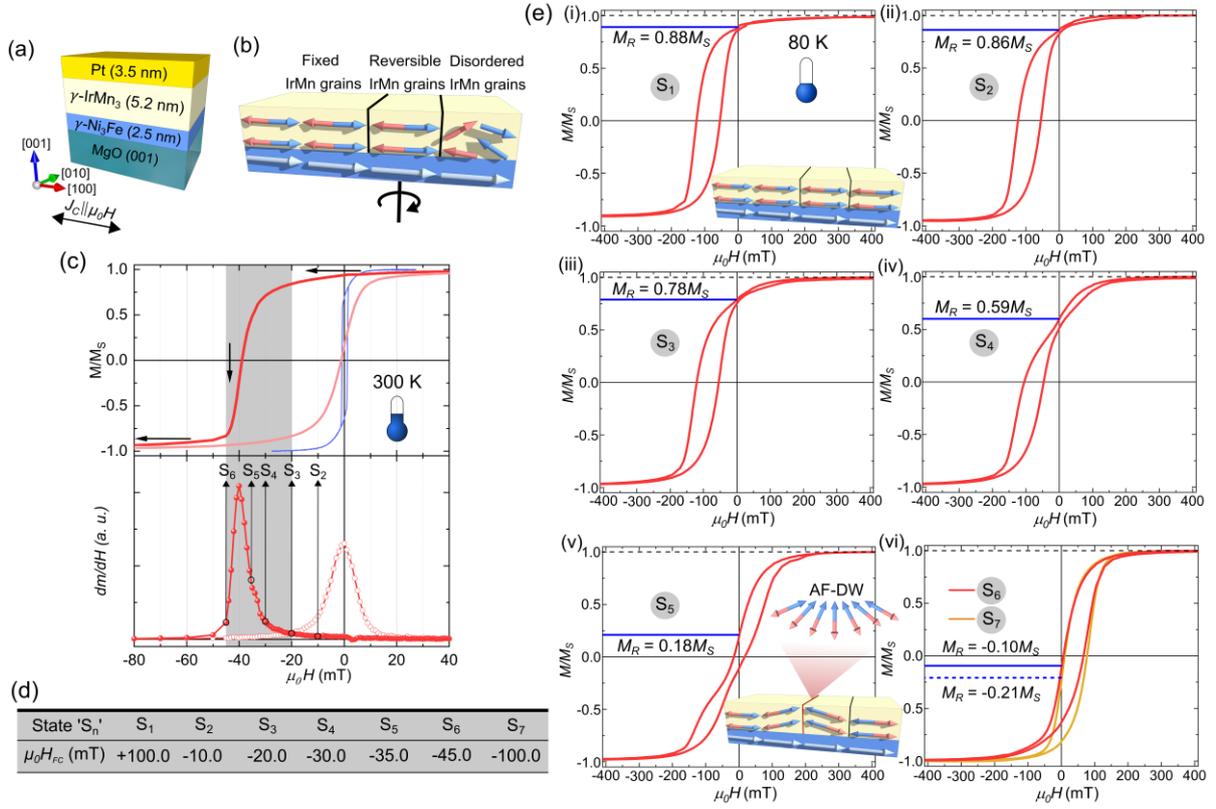

**Figure 3.** (a) Sample structure: MgO(001)/Ni₃Fe(2.5 nm)/IrMn₃ (5.2 nm)/Pt(3.5 nm). (b) Simplified spin structure of the FM/AFM heterostructure at 300 K after the field annealing process. (c) EB hysteresis loop and the corresponding first derivative. The magnetic hysteresis loop of the FM Ni₃Fe(2.5 nm) thin film without an AFM layer is shown in blue. (d) Look-up table with state number 'Sₙ' (n = 1 to 7) and the corresponding $H_{FC}$, formulated from the descending branch of the EB hysteresis loop shown in (c). (e)(i-vi) EB hysteresis loops obtained in various MFC states at 80 K using the iterative MFC steps 3 and 4 shown in Figure 1(a). The change in the remanent magnetization $M_R$ in various MFC states are indicated with solid blue lines. The insets in e(i) and e(v) show the simplified FM/AFM spin structure (at 80 K). **Note**: all the EB hysteresis loops were obtained after the EB training procedure described in the experimental section.

The reference magnetic hysteresis loop of an unbiased single FM Ni₃Fe(2.5 nm) thin film (without the AFM ) is shown in Figure 3(c) (solid blue line). We formulate a look-up table from the descending branch of the EB hysteresis loop shown in Figure 3(c). We allocate the magnetic field to the state number Sₙ for convenience, as shown in Figure 3(d). We now cool the sample from 300 K down to 80 K in $\mu_0 H_{FC} = +100$ mT (state S₁), parallel to the initial MFC direction (+7 T). Figure 3(e)(i) shows the characteristic EB hysteresis loop obtained at 80 K after the EB training process. In the MFC state S₁ ($\mu_0 H_{FC} = +100$ mT), the FM Ni₃Fe(2.5 nm) thin film is fully saturated during the cooling process (from 300 K to 80 K) in step 3(see Figure 1(a)). The disordered AFM (IrMn₃) grains develop spontaneous AFM order with the Néel vector orientation parallel with respect to the irreversible AFM Mn moments, as shown in the inset of Figure 3(e)(i). Consequently, at 80 K, we obtain a maximum $\mu_0 H_{eb} = -89.57$ mT, $\mu_0 H_C = 33.25$ mT, $J_{eb} = 155.62(\pm 12.45)$ μJ/m² and $K_{AFM} = 25.01(\pm 2.10)$ kJ/m³ considering $d_{AFM}^{cr} =$ 2.2(±0.2) nm. Note that the critical AFM (IrMn$_3$) film thickness needed to establish an EB field is significantly less at 80 K than 300 K.[26-28]

Later, we perform a *reset* operation to switch from the MFC state S$_1$ to the initial state by simply heating the sample back to 300 K in a zero magnetic field. Similarly, we iteratively repeat steps 3 and 4 (see Figure 1(a)) for all the remaining states by applying the corresponding $H_{FC}$ (with opposite sign) as listed in the look-up table shown in Figure 3(d). Figures 3e(ii-vi) show the EB hysteresis loops for all the remaining MFC states (S$_2$ to S$_7$) obtained after the EB training process. For instance, in the MFC states S$_3$ to S$_6$ (near the FM switching field at 300 K), the FM Ni$_3$Fe(2.5 nm) layer remains in a multi-domain configuration. Field cooling the AFM layer with the FM layer remaining in the multi-domain configuration results in a fraction of the AFM Mn-moments corresponding to the thermally disordered and reversible AFM grains to align anti-parallel with respect to the irreversible AFM Mn-moments. As a result, the FM Ni$_3$Fe(2.5 nm) magnetization is pinned in both positive and negative bias field directions (see Figures 3(e)(iii-vi)); equation 1 can be applied here to estimate the EB field. In particular, Figures 3 (e)(iv) and (e)(v) show bifurcated loops, which indicate the parallel impact of positively and negatively exchange biased sample regions.[29]

Assuming a multi-domain configuration of the AFM thin film in MFC states S$_3$ to S$_7$ (at 80 K), Malozemoff's random field model[66] can be applied; where perpendicular AFM domain walls are expected to nucleate during the MFC process (in *step 3*) to separate the AFM domains with opposite Néel vectors (see inset sketch in Figure 3(v)).[18,25,66] We estimate the characteristic length scale of the frozen-in AFM domains responsible for pinning the FM layer at 80 K to be $l_{\text{AFM}} \cong \pi \sqrt{A_{AFM}/K_{AFM}} \approx 45(\pm 2)$ nm,[66] using $A_{AFM}$ = 5.13 x $10^{-12}$ J/m and $K_{AFM}$ = 25.01(±2.10) kJ/m$^3$ determined earlier. In AFM IrMn$_3$(5.2 nm) thin films, planar domain walls parallel to the FM/AFM interface,[25,65] or FM/AFM Mauri-type exchange springs are unlikely to occur.[67] Moreover, a significant reduction in the exchange stiffness and intergranular exchange coupling in granular AFM thin films can also result in isolated AFM single domains.[53,62]

Figure 4(a) shows $H_{eb}$ and $H_C$ for various MFC states (from S$_1$ to S$_7$) at $T_{meas}$ = 80 K obtained from the EB hysteresis loops shown in Figure 3(e). The $H_{eb}$ dependence closely follows a sigmoid function. If we were to assume a fully reversible AFM order for the IrMn$_3$(5.2 nm) thin film during the MFC process (from 300 K to 80 K), the magnitude of $H_{eb}$ in MFC states S$_1$ and S$_7$ should be equal. However, this is not the case here. The sign change of $H_{eb}$ occurs between MFC states S$_3$ to S$_6$, closely following the FM hysteresis distribution (or the FM domain state) at 300 K, as indicated by the grey highlighted region in Figures 3(c) and 4(a). Additionally, the tunable $M_R$ and $H_{eb}$ observed in Figures 3(e) and 4(a) indicate a highly responsive MFC-dependent magnetization reversal behavior of the exchange coupled Ni$_3$Fe(2.5 nm)/IrMn$_3$(5.2 nm) heterostructures.

We performed magnetotransport measurements on the same sample with a four-point probe technique in the current-in-plane (CIP) geometry.[68] The magnetotransport measurements were performed at $T_{meas}$ = 80 K in four different MFC states while applying a non-perturbative probe current $I$ = 4 mA ($J_C \parallel \mu_0 H \parallel$[100]). The inset in Figure 4(b) shows the measurement geometry. Figure 4(b) shows the magnetoresistance (MR) $R_{4p}(\mu_0 H)$ loops for the FM/AFM heterostructure in the MFC states S$_1$, S$_4$, S$_5$, and S$_7$. The high resistivity of the IrMn$_3$(5.2 nm) thin film ($\rho_{IrMn} \sim$ 232 μΩ·cm) due to the crystal grain boundaries or non-periodic impurities and surface scattering, resulting in high spin-flip scattering and a short spin-diffusion length λ<1 nm of the conduction electrons; it was not possible to detect the magnetization reversal of the FM Ni$_3$Fe(2.5 nm) thin film in the CIP geometry. Hence, we did not observe characteristic negative or positive dips in the measured MR signal, typically expected for the FM reversal.[69] Above ±300 mT, the magnetization of the FM Ni$_3$Fe(2.5 nm) layer is fully saturated (see Figure 3(e)). The-refore, the FM domain contribution to $R_{sat}$ (at $\mu_0 H >300$ mT) can be ruled.

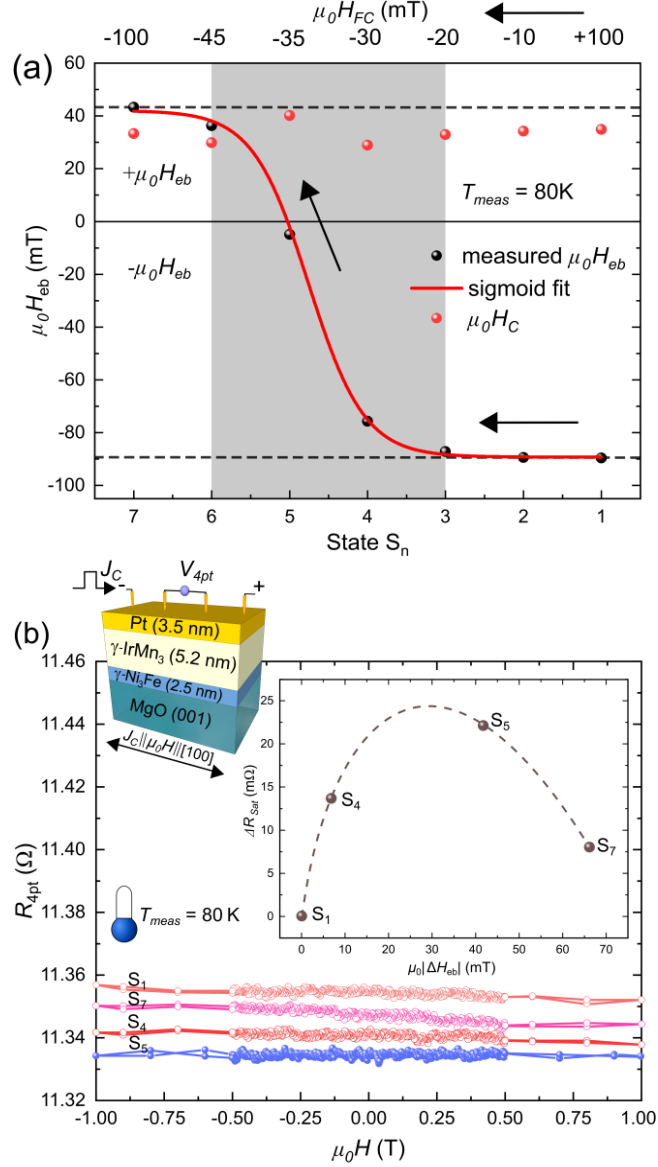

**Figure 4. (a)** The exchange bias field calculated from the series of EB hysteresis loops shown in Figure 3(e) for various MFC states (S$_1$-S$_7$) shows an apparent change in the sign and magnitude of $H_{eb}$. The highlighted region in light grey color indicates MFC states S$_3$ to S$_6$, where the sign change of $H_{eb}$ occurs. **(b)** Magnetoresistive hysteresis loops R($\mu_0 H$) with $J_C$∥[100], $\mu_0$H∥[100] in various MFC states at 80 K. The first inset shows the sample structure and the four-point probe measurement geometry. The second inset shows the correlation between $\Delta R_{sat}$ and $\Delta H_{eb}$ in various MFC states, and dashed lines are a guide to the eye.

Typically, high magnetic field pulses (up to 60 T) are needed to manipulate the AFM Néel vector.[3] The maximum magnetic field available in our experiments is limited to 7 T for magnetic hysteresis measurements and 5 T for the magneto-transport measurements. Therefore, the applied magnetic field is inadequate to rotate the AFM Mn-sublattice magnetization. However, MFC the exchange coupled FM/AFM heterostructures in various magnetic domain states of the FM can be employed to modify the AFM domain configuration.[22] In the MFC state $S_1$, when the AFM Mn-moments are parallel to the applied current direction, we achieve $R_{sat}$ = 11.3562 Ω. For instance, in the MFC state $S_4$, we expect the AFM Mn-moments corresponding to a portion of disordered and reversible AFM grains to align anti-parallel with respect to the irreversible (frozen-in) AFM Mn-moments. Consequently, we observe a decrease in $R_{sat}$, with a change in saturation resistance $\Delta R_{sat} = R_{sat}(S_1)\text{-}R_{sat}(S_4)$ of 13.7 mΩ. Similarly, we obtain $\Delta R_{sat}$ = 22.2 mΩ and 8 mΩ in states $S_5$ and $S_7$, respectively. The inset in Figure 4(b) shows the correlation between $\Delta R_{sat}$ and the change in EB field $|\Delta H_{eb}| = 0.5\times|H_{eb}(S_1)\text{-}H_{eb}(S_n)|$ in various MFC states. $\Delta R_{sat}$ reached a maximum in the multi-domain cooling states $S_5$. Moreover, $\Delta R_{sat}$ = 8….22.2 mΩ measured in various MFC states corresponding to the microscopic restructuring of the AFM domain structure compares well with previous experimental results reported in the literature.[17,18]

Due to the AFM collinear spin structure and weak spin-orbit coupling (SOC) expected in γ-IrMn₃ alloy thin films, a large MR effect has not been reported.[49-51] In chemically disordered γ-IrMn₃ alloys, the atomic occupancy probability of Mn-atoms to occupy the face-centered positions and Ir-atoms to occupy the cubic corners of the fcc unit cell depends on the stoichiometry (it is most likely statistically disordered, see Figure 5(d)).[51] Rutherford backscattering spectroscopy (RBS) measurements of the reference thin films revealed the atomic stoichiometry of the $Ir_xMn_{100-x}$ alloy with $x$ = 30($\pm$1) at. %. The Mn-atoms randomly occupying the Ir sites can deter the SOC and modify the electronic density of states (DOS).[49,50] In addition to the chemical disorder, IrMn₃ thin films studied here are tetragonally distorted ($c \neq a$). Earlier theoretical studies on IrMn alloys reported that the distance between the Mn-atoms alone can significantly influence the sign and magnitude of the Heisenberg exchange interaction and modify the electronic properties.[47] Moreover, large MR effects were predicted only for the chemically ordered $L1_2$-IrMn₃ alloys with non-collinear (triangular or $T1$-type)[49] AFM spin structure.[50] The weak magnetoresistive effects in AFM γ-IrMn₃ alloys can be attributed to the reduced SOC combined with the chemical and structural disorder.[7,49,50] Nevertheless, by manipulating the disordered and reversible AFM Mn-moments in exchange coupled Ni₃Fe/IrMn₃ heterostructures, we achieve tunable magnetic ($H_{eb}$ and $M_R$) and magnetotransport properties in various MFC states.

## 2.4. Structural Properties of MgO(001)/NiFe₃/IrMn₃/ NiFe₃/CoO Heterostructures.

In our attempt to enhance the MR ratio and further understand the role of integrating γ-IrMn₃ thin film with coexisting AFM phases into a complex exchange coupled heterostructured system, we consider a MgO(001)/γ-Ni₃Fe(2.5nm)/γ-IrMn₃(5.2nm)/γ-Ni₃Fe(15nm) /CoO(2.0 nm) heterostructure (see Figure 5(a)). Figures 5(b) -5(d) show the primitive unit cells (lattice and spin structure) of CoO, γ-Ni₃Fe, and γ-IrMn₃ alloys. We performed X-ray photoelectron spectroscopy (XPS) to study the oxidation states of Co. The Co2$p$ spectrum shown in Figure 5(e) reveals an observable satellite feature at ~785.3 eV corresponding to the Co(II) oxide state. The absence of an additional peak at ~778.2 eV corresponding to the metallic Co indicates complete thin film oxidation.

Additionally, we performed structural characterization using XRD. Please note that the samples were annealed at 605 K in a +7 T magnetic field for 90 min, and subsequently, MFC down to 300 K before performing the structural characterization. Figure 5(f) shows a $\theta$-$2\theta$ scan along the [001] orientation of the single-crystal MgO substrate. The fundamental (002) diffraction peaks corresponding to the CoO, IrMn₃, Ni₃Fe thin films, and the single-crystal MgO substrate are labeled. We calculate the $c$-axis lattice distance from the diffraction peak positions of CoO ($c$ = 4.44 Å), γ-IrMn₃ ($c$ = 3.69 Å), and γ-Ni₃Fe ($c$ = 3.42 Å). We use the Scherrer equation to estimate the structural coherence length $\tau$ along the growth direction for CoO ($d = \tau = 2.00(\pm0.20)$ nm), IrMn$_3$ ($d = \tau = 5.20(\pm0.30)$ nm), and Ni$_3$Fe ($d \neq \tau = 7.34(\pm0.35)$ nm) thin films; where $d$ is the film thickness. The rocking ($\omega$) scans along the fundamental IrMn$_3$ (002) and Ni$_3$Fe (002) Bragg peaks reveal a rather narrow rocking width of ~0.11° and ~0.10°, respectively (see inset in Figure 5(f)). The $\phi$-scans along (111) reflections of the substrate confirm the epitaxial orientation relationship between the single-crystal MgO substrate and the Ni$_3$Fe(IrMn$_3$) thin films with $\Delta\phi \approx$ 6.31°(8.81°), MgO(001)[100]$_{fcc}$‖Ni$_3$Fe(001)[100]$_{fct}$‖IrMn$_3$ (001)[100]$_{fcc}$‖Ni$_3$Fe(001)[100]$_{fct}$. Additionally, the impact of field annealing on the magnetization of the Ni$_3$Fe(15 nm) thin film is shown in Supplementary Figure S1.

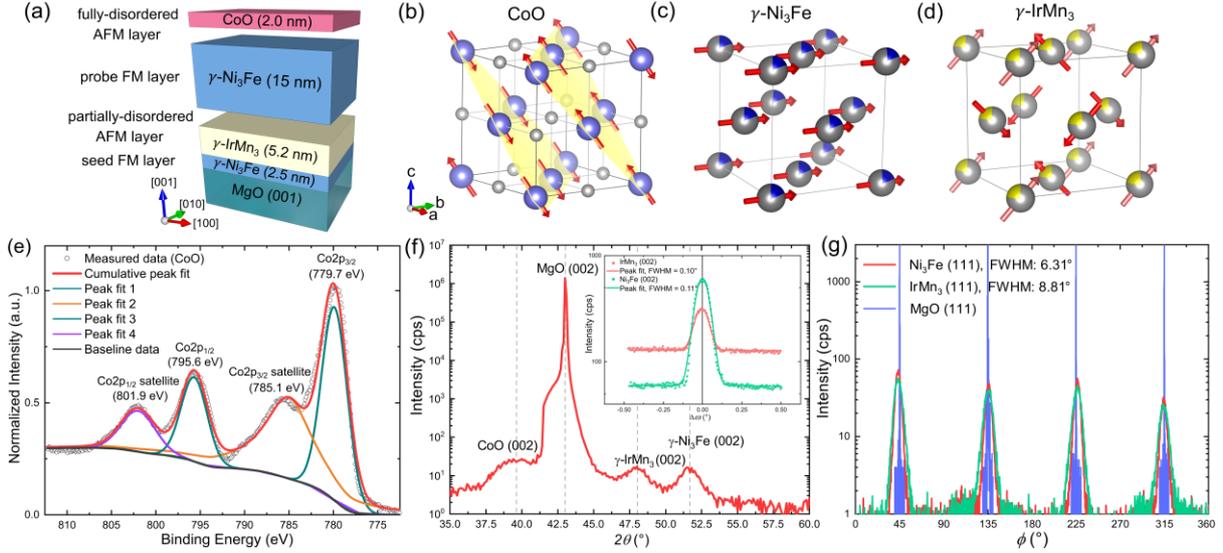

**Figure 5. (a)** Sample structure of the MgO(001)/Ni3Fe(2.5 nm)/IrMn3(5.2 nm)/Ni3Fe(15 nm)/CoO(2.0 nm) sample. **(b)** NaCl (rock salt) type crystal structure of CoO. Below TN, the magnetic moments of Co$^{2+}$ ions are aligned parallel (FM) within the {111} plane due to the direct FM interaction. In contrast, an anti-parallel alignment between the adjacent {111} planes can be seen due to the AFM superexchange between Co$^{2+}$-O$^{2-}$-Co$^{2+}$ ions.[70,71] **(c and d)** Chemically disordered $\gamma$-Ni$_3$Fe and $\gamma$-IrMn$_3$ fcc crystal structure. The Mn magnetic moments are tilted by ~45° from the crystal face diagonal towards the cubic faces.[51] **(e)** Selected energy XPS spectrum with the satellite features of Co2p at ~785.1 eV and ~801.9 eV confirming the Co(II) oxidation state. **(f)** XRD θ-2θ scans along the [001] orientation of the substrate. The inset shows the rocking curves for the IrMn3(5.2 nm) and Ni3Fe(15 nm) thin films at the (002) reflection. **(g)** $\phi$-scans for the MgO substrate, IrMn3, and Ni3Fe thin films along the (111) reflections.

## 2.5. Magnetization Reversal and Magnetotransport Properties of MgO(001)/NiFe3/IrMn3/NiFe3/CoO Heterostructures.

Figure 6(a) shows the layer structure and the magnetotransport measurement geometry. Figure 6(b) shows the linear I-V curves for the heterostructures with and without the ultra-thin AFM CoO insertion layer. Inserting an ultrathin CoO(2.0 nm) film resulted in a nominal increase in the measured resistance $\Delta R_{4pt} = 1.1768$ $\Omega$ at 150 K compared to the heterostructure without the CoO layer.[72] The measured I-V curve did not reveal any non-linear characteristics, and the applied current is well within the linear ohmic region. Therefore, ionic transport can be excluded. Additionally, we deposited a Pt(3.5 nm) cap layer to protect the exchange coupled heterostructures from oxidation. After the sputter deposition, the samples were field annealed at 605 K for 90 min. During the sputtering and annealing process, Pt atoms can inter-diffuse into the grain boundaries of the CoO layer.[57,73] They may also occupy the interstitial lattice sites and/or vacancy defects within the CoO thin film. The interstitial Pt atoms not only increase the density of the CoO thin film and modify the electronic properties, but they can also promote the electrical conductivity (via electron transport), *i.e.*, cause a substantial decrease in the resistance of the ultra-thin CoO layer.[72]

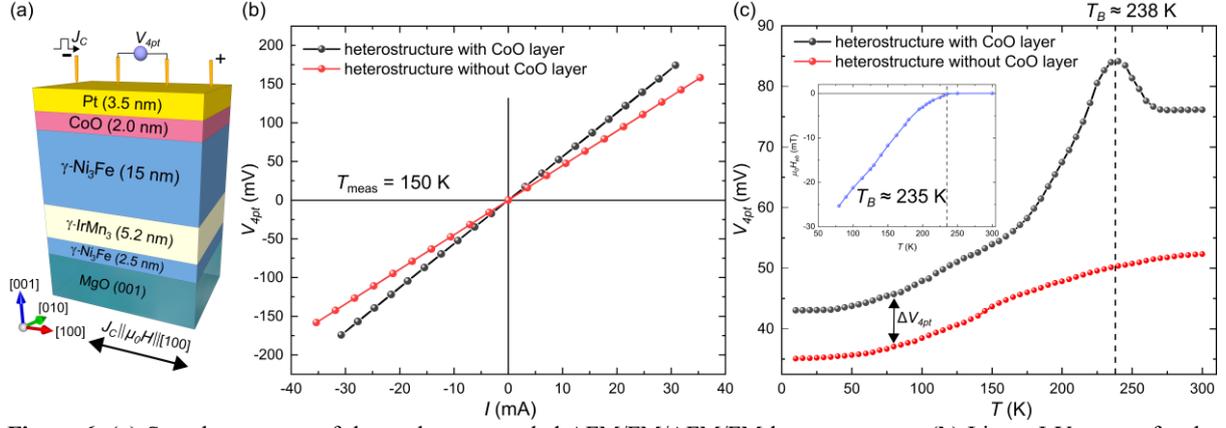

**Figure 6. (a)** Sample structure of the exchange coupled AFM/FM/AFM/FM heterostructure. **(b)** Linear I-V curves for the sample with and without the ultra-thin AFM CoO layer. **(c)** Voltage drop V4pt as a function of measurement temperature at $I$ = 10 mA ($J_C$||[100]). The inset shows $H_{eb}(T)$ for the reference bilayer sample, MgO(001)/Ni$_3$Fe(15 nm)/CoO(2.0 nm)/Pt(3.5 nm).

Figure 6(c) shows the voltage drop as a function of the measurement temperature $V_{4pt}$(T) for the exchange coupled heterostructures with and without the CoO insertion layer. As the CoO thin film undergoes a phase transition from a paramag- netic to an AFM state; we found a significant increase in the voltage drop at ~238 K ($\ll T_N$ of AFM bulk CoO). In contrast, the sample without the ultrathin CoO film did not reveal any characteristic peak below 300 K. We determine $T_B \approx 235$ K for the exchange coupled Ni$_3$Fe(15 nm)/CoO(2.0 nm) interface from the reference bilayer sample. The inset in At the top Ni$_3$Fe/CoO interface, as the AFM CoO grains undergo a phase transition from a paramagnetic to an AFM state, the FM Ni$_3$Fe(15 nm) domains experience sporadic magnetic switching events, most likely forming fractal multi-domains. As a result, the spin-polarized charge carriers (electrons) experience significant scattering at the domain boundaries,[74-77] increasing the measured voltage drop near $T_B$. Generally, a broad peak indicates a considerable distribution of the ordering temperature within the granular AFM CoO thin film.

Figure 7(a) shows the EB hysteresis loop measured at 300 K after the initial field annealing and EB training process for the exchange coupled MgO(001)/Ni$_3$Fe(2.5nm)/IrMn$_3$(5.2 nm)/Ni$_3$Fe(15nm)/CoO(0 and 2.0 nm)/Pt(3.5 nm) heterostructures. At 300 K, CoO remains in a paramagnetic state.[70] Therefore, EB hysteresis loops for the heterostructures with and without the CoO(2.0 nm) insertion layer are similar. The magnetization reversal of the top FM Ni$_3$Fe(15 nm) layer occurs via sporadic nucleation of fractal magnetic domains and an increase in the population of magnetic domains with opposite magnetization, followed by the saturation of the FM. The scanning Kerr microscopy (SKerrM) images corresponding to the magnetization reversal of the top Ni$_3$Fe(15 nm) layer closely following the descending branch of the EB hysteresis loop at 300 K are shown as insets in Figures 7(b)(i)-(b)(v). The evolution of the magnetic domain structure and tetragonality ($c/a$-ratio) in the FM Ni$_3$Fe(15 nm) layer as a function of AFM IrMn$_3$ film thickness $d_{IrMn3}$ are shown in Supplementary Figures S2 and S3, respectively. Due to the decrease in the tetragonality ($c/a$-ratio) of the FM $Ni_3Fe$(15 nm) thin films with AFM $IrMn_3$ film thickness >10 nm, we found a significant increase in the domain size from a micrometer size fractal magnetic domain structure to millimeter size domains with somewhat smooth domain wall motion. Further, we obtain an anisotropic magnetoresistance (AMR) ratio of 0.98% for the exchange coupled heterostructure at 300 K (see Supplementary Figure S4).[78] The significant decrease in the AMR ratio of the exchange coupled heterostructures observed at 300 K, in contrast to the typical bulk values (4-5%) reported for $Ni_3Fe$ alloys,[78] can be attributed to the extrinsic contributions, namely grain boundary or impurity scattering, surface and interfacial scattering, and strain-induced effects.

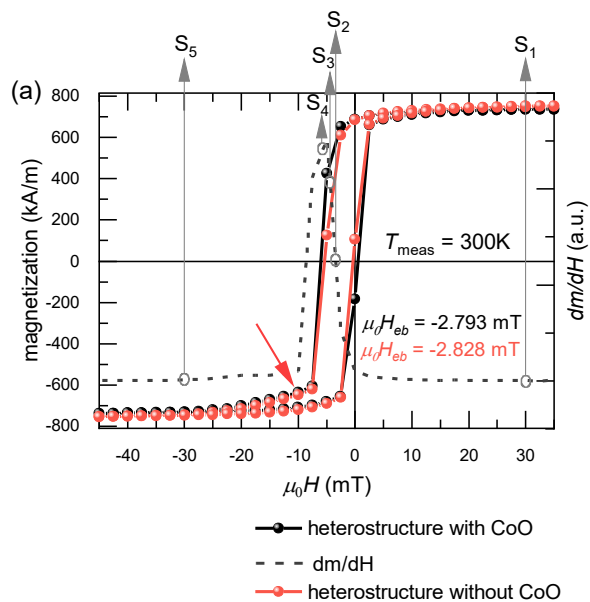

**Figure 7. (a)** The EB hysteresis loop measured at 300 K after EB training process for the exchange coupled MgO(001)/$Ni_3Fe$(2.5 nm)/$IrMn_3$(5.2 nm)/$Ni_3Fe$(15 nm)/CoO(0 and 2.0 nm)/Pt(3.5 nm) heterostructures. The states $S_1$-$S_5$ highlights the FM domain states and the corresponding $H_{FC}$, and the red arrow indicates the contribution of the FM $Ni_3Fe$(2.5 nm) thin film.

The red arrow in Figure 7(a) indicates the bottom FM $Ni_3Fe$(2.5 nm) layer contribution to the EB hysteresis loop. Additionally, we performed MOKE magnetometry on the exchange coupled heterostructures with optical access from both the front and the backside of the sample (see Supplementary Figures S5 (b) and (c)). The MOKE hysteresis loops obtained at 300 K from the backside of the sample further confirm the presence of the highlighted tail feature corresponding to the bottom FM $Ni_3Fe$(2.5 nm) thin film. In exchange coupled films with $IrMn_3$(5.2 nm), assuming a full structural coherence length of the AFM grains along the growth direction, we expect a bulk AFM spin structure-dependent magnetization reversal in the adjacent FM thin films.[79]

Figure 7(b) shows the magnetization and MR hysteresis loops in various MFC states. Starting from 300 K, we cool the samples down to 80 K in the saturation state of the FM, $\mu_0 H_{FC}$ = +30 mT ($S_1$). The magnetic contrast corresponding to the top $Ni_3Fe$(15 nm) layer at 300 K in the MFC state $S_1$ is shown as an inset in Figure 7(b)(i) and indicates the complete saturation of the FM. The bottom $IrMn_3$/$Ni_3Fe$(15 nm) and the top $Ni_3Fe$(15 nm)/CoO interfaces are field cooled in a positive saturation state of the FM. Consequently, we obtain a negative shift in the EB hysteresis loops, as shown in Figure 7(b)(i).

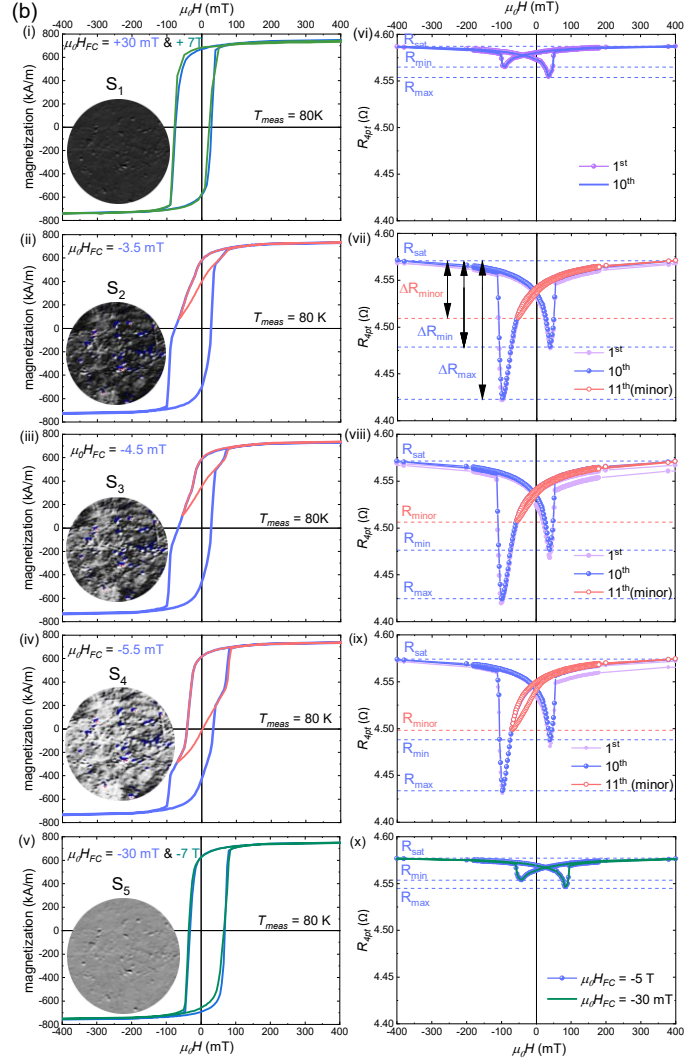

**Figure 7. (b)** Magnetic and magnetic transport properties of the exchange coupled heterostructure **(i-v)** EB hysteresis loops obtained at 80 K in various MFC states ($S_1$-$S_5$). The insets show the SKerrM images of the sample obtained at 300 K at various applied magnetic field values closely following the descending branch of the EB hysteresis loop shown in Figure 7(a). **(vi-x)** Magnetoresistance curves $R_{4pt}(\mu_0 H)$ obtained at 80 K in various MFC states ($S_1$ to $S_5$). **Note**: $R_{sat}$, $R_{max}$, $R_{min}$, and $R_{minor}$ are the saturation resistance, maximum resistance, minimum resistance, and the resistance obtained from the minor loops, respectively. The change in resistance, $\Delta R_{max} = R_{sat}$-$R_{max}$, $\Delta R_{min} = R_{sat}$-$R_{min}$, and $\Delta R_{minor} = R_{sat}$-$R_{minor}$.

The fully disordered and reversible AFM-moments at the top $Ni_3Fe$(15 nm)/CoO and the bottom $IrMn_3$/$Ni_3Fe$(15 nm) interfaces tend to align parallel with respect to the irreversible AFM Mn-moments. As a result, we do not expect perpendicular domain walls to form in either the FM or AFM layers during (or after) the MFC process. Similarly, the MR loop reveals a shift in the negative magnetic field direction (see Figure 7(b)(vi)). For $J_C \| \mu_0 H$, we observe a negative change in resistance as a function of the applied magnetic field.[75,80] The negative resistance dips occur near the coercive field of the top $Ni_3Fe$(15 nm) layer, revealing a domain wall-driven change in resistance.[69,75,80] Further increase in the applied magnetic field results in the saturation of the FM. We obtain a MR ratio of $\Delta R/R = (R_{sat}$-$R_{max})/R_{max}$ = 0.73%. Hence, the FM contribution to MR is also saturated. As shown earlier in Figure 4(b), due to the high resistivity and short spin-diffusion length of the IrMn$_3$(5.2 nm) thin film, we did not detect the magnetization reversal of the bottom FM $\gamma$-Ni$_3$Fe(2.5 nm) layer.

After performing a *reset* operation, the sample is MFC in $\mu_0 H_{FC}$ = -3.5 mT (S$_2$). The magnetic contrast of the top $\gamma$-Ni$_3$Fe(15 nm) layer in the MFC state S$_2$ at 300 K is shown as an inset in Figure 7(b)(ii). It indicates a multi-domain state with small fractal magnetic domains. The AFM Néel vector orientation of the fully disordered and reversible AFM grains at the top Ni$_3$Fe(15 nm)/CoO and the bottom IrMn$_3$/Ni$_3$Fe(15 nm) interfaces strongly depends on the domain state of the FM grains in proximity during the MFC process. Note that the irreversible AFM Mn-moments retain their preset spin orientation (Néel vector). Therefore, after the MFC process, a fraction of the AFM moments corresponding to the disordered and reversible AFM grains at the top and the bottom (FM/AFM) interfaces tend to align either parallel or anti-parallel with respect to the irreversible AFM Mn-moments. Consequently, the top and the bottom AFM layers are expected to be in a multi-domain state; this essentially gives rise to the bimodal, partially positive, and partially negative shift in the same magnetic hysteresis loop. The MOKE hysteresis loops measured at various temperatures for the exchange coupled heterostructures obtained in various MFC states (see supplementary Figures S5 and S6) further confirm the characteristic bimodal EB behavior observed in Figures 7(b)(i)-(b)(v). At 80 K, when the sample is MFC in states S$_2$, S$_3$, and S$_4$, a portion of the FM that is exchange biased in the positive magnetic field direction can be switched via minor EB and MR hysteresis loops (see Figures 7(b)(ii-iv) and 7(b)(vii-ix)). The MR contribution due to the positively biased part of the FM can be extracted from the minor MR loops ($\Delta R_{minor}$) shown in Figure 7(b)(vii-ix)). In all three cases (states S$_2$-S$_4$), the minor magnetization and MR hysteresis loops closely follow the full loop. This type of magnetization reversal mechanism indicates that the fraction of the FM that is exchange biased along the positive magnetic field direction can be switched independently without influencing the negative exchange biased portion of the FM magnetization. Néel-type domain walls perpendicular to the interface plane are expected to form in both FM as well as the AFM layers to separate the magnetic (or AFM) domains with opposite spin orientation. We suggest that such perpendicular domain walls in the exchange coupled heterostructures are the primary source of the observed 4-5 fold increase in the MR ratio in states S$_2$-S$_4$.[74-77,80] The change in resistance of the sample depending on the domain wall width $\delta_w$ and domain size $\delta_S$ is given by $\Delta R/R \propto (2p/(1-p)^2)/\delta_w \delta_S$,[75] where $p$ is the ratio of the mean free path of spin-up and spin-down electrons. In various MFC states, we do not expect $\delta_w$ to change significantly. However, in multi-domain MFC states, namely S$_2$, S$_3$, and S$_4$, we observe a significant increase in the population of fractal magnetic domains separated by the domain walls. Therefore, the spin-polarized conduction electrons undergo multiple scattering events at the domain boundaries causing a sizable enhancement in MR ratio (up to 3.47% in the MFC state S$_2$, see Figure 7(b)(vii)).[74-77]

In the MFC state S$_5$, the FM layer is fully saturated in the negative field direction (see inset in Figure 7(b)(v)). Most of the fully disordered and reversible AFM moments at the top and the bottom interfaces tend to align anti-parallel with respect to the irreversible AFM Mn-moments. As a result, the magnetic and magnetoresistive hysteresis loops are shifted to the positive magnetic field direction (see Figures 7(b)(v and x)). In MFC state S$_5$, we do not expect a significant amount of domain walls perpendicular to the FM/AFM interfaces to nucleate and stabilize at 80 K. Consequently, we see a drastic reduction in the MR ratio from 3.14% (in state S$_4$) to 0.72% (in state S$_5$). We achieve tunable multi-level MR states by systematically manipulating and stabilizing the magnetic (and AFM) domain population in the exchange coupled heterostructures in various MFC states.

In the positive and negative bias field direction of the MR loops shown in Figures 7(b)(v-x), we can see an asymmetric change in the resistance.[80] In particular, the difference in MR for the MFC states $S_2$, $S_3$, and $S_4$ in the negative bias field direction ($\Delta R_{max}$) is much larger than the resistance change observed in the positive bias field direction ($\Delta R_{min}$). Similarly, the asymmetric magnetization reversal can be observed in the EB hysteresis loops as well (see Figures 7(b)(ii-iv)). This behavior can be attributed to the asymmetric magnetic domain formation typically observed in the exchange coupled FM/AFM systems.[33,34] Hence, we suggest that the asymmetric magnetic domain formation is the most likely reason for the observed asymmetries in the magnetic and MR hysteresis loops in various MFC states. Due to the small fractal magnetic domains with sporadic nucleation events observed in the exchange coupled heterostructures with IrMn$_3$(5.2 nm), we could not capture the asymmetric magnetic domain formation along the biased and the unbiased branches of the EB hysteresis loop. Nevertheless, in samples with IrMn$_3$(16 nm), we found conclusive evidence for the asymmetric magnetic domain formation (see Supplementary Figure S9(e)).

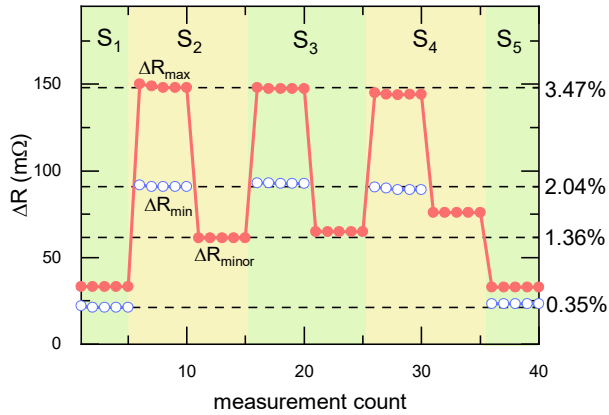

**Figure 8.** Multi-level magnetoresistance states measured at 80 K, obtained using the iterative MFC process.

The MR measurements were repeated to investigate the interfacial EB training-related jitter. Figure 8 shows the measured $\Delta R$ for various MFC states ($S_1$-$S_5$). We found a minor change in $\Delta R$ as a function of measurement count for a given state. The second resistance levels $\Delta R_{minor}$ in each MFC state are obtained from the minor loops (see MR minor loops in Figures 7(b)(vii-xi)). By selecting a pre-formulated write field $H_{FC}$ sequence (see Figure 7(a) states $S_1$-$S_5$), one can attain complex MR states ranging from simple binary, cumulative, or even random resistance states.

Figure 9(a) shows the measured $H_{eb}$ as a function of $H_{FC}$ at 80 K for the exchange coupled heterostructures with and without the AFM CoO insertion layer. A comprehensive study of the magnetic, MOKE, and magnetotransport properties of the reference exchange coupled heterostructures without the CoO insertion layer are presented in Supplementary Figures S7-S8. In the regime with IrMn$_3$(5.2 nm), the domain state of the FM layer during the iterative MFC process (steps 3 and 4 in Figure 1(a)) can be used to achieve a tunable $H_{eb}$, similar to the magnetization reversal behavior observed for the exchange coupled Ni$_3$Fe(2.5 nm)/IrMn$_3$(5.2 nm) heterostructure (see Figure 4(a)). Note that the sign change of $H_{eb}$ from (initially set) negative to positive bias field direction occurs gradually following the FMs domain state during the iterative MFC process (additionally, see Figures 7(a) and 8(i-v)). The magnitude of $H_{eb}$ is consistently higher for heterostructures with the AFM CoO(2.0 nm) layer due to the additional bias field at the top Ni$_3$Fe(15 nm)/CoO(2.0 nm) interface.

In the latter regime with IrMn$_3$(16 nm), we address the impact of increasing the IrMn$_3$ grain volume on the MFC-dependent magnetization reversal properties of the exchange coupled heterostructures. The magnetic and magnetotransport properties of the exchange coupled heterostructures with IrMn$_3$(16 nm) in various MFC states are presented in Supplementary Figures S9 and S10. Figure 9(b) shows $H_{eb}$ as a function of $H_{FC}$ at 80 K. For the heterostructures without the AFM CoO insertion layer, $H_{eb}$ remains in the negative bias field direction. The stability of the EB field strength in this regime can be attributed to the increase in $\Delta E$ of the AFM IrMn$_3$ grains.[26] Due to the persistence of thermally disordered and reversible AFM moments, we found a slight reduction in EB field strength when the samples were MFC in the negative magnetic field direction. Introducing an ultrathin AFM CoO(2.0 nm) film with a fully reversible Néel order, as shown in Figure 9(b), resulted in a certain degree of tunability in the system. The change in the sign and magnitude of $H_{eb}$ occurs sharply at a critical $\mu_0 H_{FC}$ = -20 mT, shown as an inset in Figure 9(b).

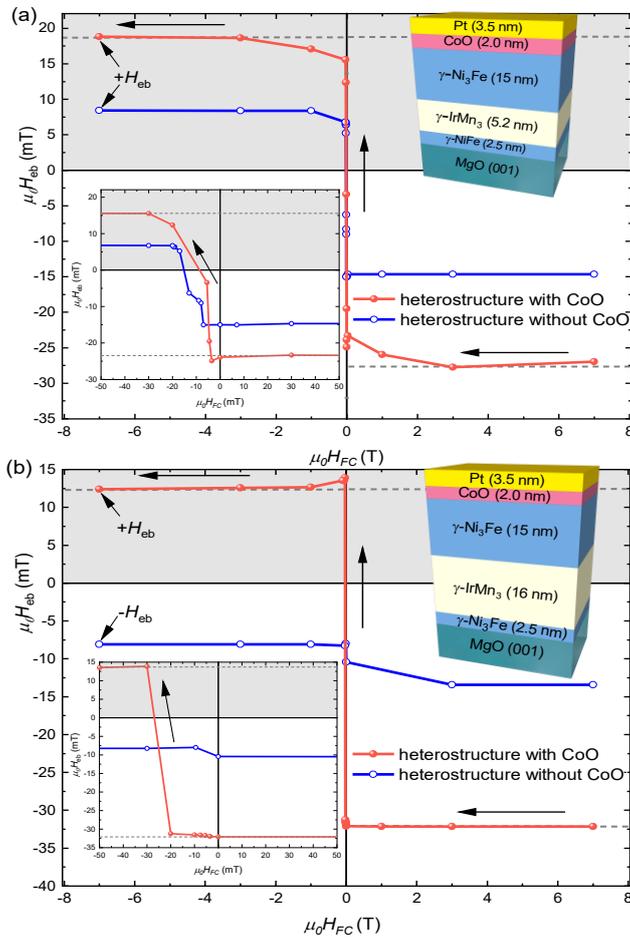

**Figure 9**. $H_{eb}$ as a function of the applied cooling field $H_{FC}$ at 80 K. **(a)** Regime I: for MgO(001)/Ni$_3$Fe(2.5nm)/IrMn$_3$(5.2nm)/Ni$_3$Fe (15 nm)/CoO(0 and 2.0 nm)/Pt(3.5 nm) heterostructures, and **(b)** Regime II: MgO(001)/Ni$_3$Fe(2.5nm)/IrMn$_3$(16nm)/Ni$_3$Fe(15nm)/CoO(0 and 2.0 nm)/Pt(3.5 nm) heterostructures.

Supplementary Figures S9 and S10. Figure 9(b) shows $H_{eb}$ as a function of $H_{FC}$ at 80 K. For the heterostructures without the AFM CoO insertion layer, $H_{eb}$ remains in the negative bias field direction. The stability of the EB field strength in this regime can be attributed to the increase in $\Delta E$ of the AFM IrMn$_3$ grains.[26] Due to the persistence of thermally disordered and reversible AFM moments, we found a slight reduction in EB field strength when the samples were MFC in the negative magnetic field direction. Introducing an ultrathin AFM CoO(2.0 nm) film with a fully reversible Néel order, as shown in Figure 9(b), resulted in a certain degree of tunability in the system. The change in the sign and magnitude of $H_{eb}$ occurs sharply at a critical $\mu_0 H_{FC}$ = -20 mT, shown as an inset in Figure 9(b).

Moreover, compared to regime I, the MR ratio ($\Delta R/R$) and the number of magnetoresistance states that can be stabilized in regime II are significantly less (see Supplementary Figure S9(c)). Therefore, we suggest a crucial ingredient for maximizing the possible number of magnetoresistance states and introducing tunable EB properties is the presence of disordered and reversible AFM moments in the FM/AFM-based heterostructures. We conclude that the observed highly tunable multi-level magnetoresistance states and EB field in FM/AFM-based heterostructures are achieved by the microscopic manipulation of the thermally disordered and reversible AFM moments in various MFC states.

To further uncover the role of disordered and reversible AFM moments in the simple exchange coupled FM/AFM heterostructures on the EB effect, we study the MFC-dependent EB properties of MgO(001)/IrMn$_3$('$d_{IrMn3}$')/Ni$_3$Fe (15 nm) heterostructures from $d_{IrMn3}$ = 0 to 35 nm (see Supplementary Figures S11, S12, and S13). We reveal that the disordered and reversible AFM moments persist even up to $d_{IrMn3}$ = 35 nm. This further confirms the inevitable existence of the AFM grains ($V<V_C$) with disordered, reversible, and frustrated AFM order in sputter-deposited AFM thin films, independent of the film thickness.

**2.6. Observations Based on Supplementary Element-specific XAS/XMCD measurements.** Despite our best efforts, various open questions remain due to the complex NiFe$_3$/IrMn$_3$/NiFe$_3$/CoO heterostructure architecture with multiple FM/AFM (AFM/FM) interfaces and elusive AFM domain structure: **a)** Does the MFC process (from 300 K to 80 K) alter the interfacial and bulk AFM spin structure? **b)** The impact of magnetic field cycling on the stability of the uncompensated Mn-moments at 80 K. **c)** Can Mauri-type planar domain walls nucleate at the IrMn$_3$/NiFe$_3$ interfaces in the granular heterostructures? [25,67]

To address the aforementioned questions, we performed supplementary synchrotron-based element-specific XAS/XMCD and magnetic hysteresis loop measurements at the AFM Mn and the FM Ni and Co $L_{2,3}$-edges on similar exchange coupled heterostructures in various MFC states (see Supplementary Figures S14-S19 and Table S1). **a)** Applying a cooling field (from 300 K to 80 K) anti-parallel with respect to the preset bias field direction of the AFM Mn-moments resulted in ~25% (~22%) reduction of $|H_{eb}|$ at the Mn/Ni (FM/CoO) interface. The significant decrease of the EB field can be attributed to the modification of the interfacial and bulk AFM spin structure[32,37,79] (see Supplementary Figures S16-S18 and Table S1). Additionally, we found a ~104% (~153%) increase in $H_C$ at the Mn/Ni (FM/CoO) interface.

**b)** The magnetic hysteresis loops obtained near the resonant Mn $L_3$-edge and XMCD ($L_{2,3}$-edges) spectra measured at $T$ = 80 K (after the EB training process) reveal excellent magnetic field stability of the uncompensated Mn-moments in both parallel and anti-parallel MFC states (see Supplementary Figure S19). **c)** Further, we found no evidence to support the existence of Mauri-type planar domain walls at the Mn/Ni interface.[67] However, AFM domain walls perpendicular to the (AFM/FM) interface can still develop, as prompted by a dramatic increase in the coercive field at the Mn/Ni interface in the anti-parallel MFC state (see Supplementary Figure S18). A summary of crucial interfacial properties of the exchange coupled heterostructures, such as coercive field, EB field, interfacial exchange energy obtained from the element-specific Ni and Co ($L_3$-edges) magnetic (XMCD) hysteresis loops are presented in Supplementary Table S1.

**3. CONCLUSIONS.** In conclusion, we explored the structural, magnetic, and magnetotransport properties of $Ni_3Fe/IrMn_3$, $Ni_3Fe/IrMn_3/$ $Ni_3Fe$, and $Ni_3Fe/IrMn_3/Ni_3Fe/CoO$ heterostructures in various MFC states. We unveil the impact of rotating the relative orientation of the disordered and reversible AFM moments with respect to the irreversible AFM moments on the magnetic and magnetoresistance properties of the exchange coupled heterostructures. Independent of the AFM film thickness (measured up to $d_{IrMn3} = 35$ nm), disordered, reversible, and frustrated AFM order persists in granular AFM $IrMn_3$ thin films, most likely caused by the inevitable structural defects and atomic site disorder. We further tuned the magnetization reversal behavior, EB field strength, and magnetoresistance states using the highly responsive (or active) MFC-dependent nature of the AFM grains with disordered and reversible AFM order in the exchange coupled heterostructures. The approach introduced in this work to systematically manipulate the disordered and reversible AFM moments and electrically detect multi-level resistance states in bimetallic and oxide AFM-based heterostructures is likely to play an essential role in designing complex resistive memory devices for future computing device applications. Our work further accentuates the strong correlation between chemical, structural, magnetic, and magnetotransport properties of the FM/bi-metallic AFM/FM/TM-oxide AFM-based heterostructures.

### 4.0. EXPERIMENTAL SECTION

**4.1. Sample preparation.** Samples investigated in this study were prepared by magnetron sputtering using Ar as sputter gas ($p_{Ar} = 3.5 \times 10^{-3}$ mbar). The double-side-polished (1 cm x 1 cm) single-crystal MgO(001) substrates were pre-annealed at 875($\pm$2) K for 5 hours to remove the thin inorganic Mg(OH) layer. Later, the substrates were cooled down to 550($\pm$2) K for sputter deposition. DC magnetron sputtering was used for all the materials except for the insulating CoO layer. Reactive RF sputtering was used to deposit the CoO layer, with $O_2$ as the reactive gas. By systematically varying the $O_2$ flow with respect to Ar gas flow in the chamber during the deposition, we were able to tune the oxidation states of Co, starting from mixed phases of metallic Co-CoO, CoO to $Co_3O_4$ (see supplementary material in ref.73). We strictly focus on the AFM CoO phase, mainly due to the high $T_N$.

Stoichiometric 2-inch alloy targets were used in the case of $Ni_{81}Fe_{19}$, $Mn_{23}Ir_{77,}$ and CoO (with 4N purity). Pt (for electrical measurements) and Al (naturally oxidized to $Al_2O_3$) were used as cap layers to protect the functional heterostructures from oxidation (or degradation). The deposition rates were pre-calibrated using X-ray reflectivity measurements. During the sputtering process, the thicknesses of the individual layers were monitored using a quartz crystal microbalance. The atomic stoichiometry of $Mn_{100-x}Ir_x$ with $x = 30(\pm1)$ at. % and $Ni_{100-x}Fe_x$ with $x = 19(\pm1)$ at. % were confirmed by RBS measurements. Due to the different sputter yields of Mn and Ir atoms, we found a significant difference between the compositions of the sputter target and the deposited thin films.

The samples for the synchrotron-based element-specific XAS/XMCD experiments were prepared at 293 K on $Si_3N_4$ (100 nm) membrane samples.

**4.2. Structural Characterization.** XRD experiments were performed using Cu–$K_\alpha$ radiation at energy 8.04 keV ($\lambda = 0.15406$ nm); we use Scherrer equation to determine the structural coherence length $\tau = K\lambda/(\Delta(2\theta)\cos\theta)$, where $\tau$ is the mean size of the ordered crystallites, $K$ is the dimensionless shape factor close to unity. We obtain $k = 2\pi/\lambda = 4.0784$ Å$^{-1}$. Note that $\Delta(2\theta)$ is the FWHM of the Bragg peaks (Ni$_3$Fe (002) and IrMn$_3$ (002)) obtained from the out-of-plane $\theta$-$2\theta$ scans. Oxidation states of Co were characterized using XPS.

Atomic force microscopy scans were performed using a NanoWizard II instrument with a (Pointprobe NCH) Si cantilever from NanoWorld AG.

**4.3. Magnetic and Magnetotransport Measurements.** Magnetic properties measurement and magnetic field annealing of the samples were performed using the magnetic properties measurement system (MPMS) superconducting quantum interference device–vibrating sample magnetometer (SQUID–VSM) with an oven option. The set-up offers a possibility to apply up to a 7 T magnetic field with a wide temperature range of 1.5–1000 K. The heating and cooling rates were kept constant at 15 K/min. All the EB hysteresis loops (at both 300 K and 80 K) were obtained after training the sample ten times in a magnetic field sequence (+7 T→ -7T→ +7T→ -7T). Magnetotransport measurements (in four-point probe geometry)[68] were performed using an Oxford cryostat with a temperature range of 4.5–300 K and a maximum magnetic field up to 5 T.

**4.4. Magneto-Optical Kerr Magnetometry and Imaging.**

SKerrM imaging and MOKE hysteresis loops in longitudinal geometry were obtained using a commercial nanoMOKE2 set-up. The magnetic imaging and hysteresis loops were obtained using a red laser with a wavelength of ~635 nm and a spot size of <10 μm, and a maximum magnetic field of 0.5 T in the film plane. The set-up is equipped with a  He flow-cryostat and a heating option with a measurement temperature range of 4.5–505 K.

**4.5. Synchrotron-based Element-Specific XAS/XMCD Measurements.** The synchrotron-based element-specific XAS/XMCD and magnetic hysteresis measurements were performed at the VEKMAG-2 end-station installed at the PM2 beamline of BESSY II. The end station offers the possibility to apply up to 9 T magnetic field in the beam direction, 2 T in the plane perpendicular to the beam direction, and 1 T in all directions using a vector coil. The set-up offers a measurement temperature range of 2–500 K and an energy range of 20–1600 eV.

**ASSOCIATED CONTENT**

**Supporting Information**

**Supplementary Figures and Tables:**

**Figure S1**. Magnetization of the FM NiFe(15 nm) layer as a function of temperature for the $\gamma$-IrMn$_3$/Ni$_3$Fe bilayers.

**Figure S2**. Evolution of the Ni$_3$Fe(15 nm) domain structure as a function of the IrMn$_3$ film thickness.

**Figure S3**. Tetragonality ($c/a$-ratio) of the Ni$_3$Fe and IrMn$_3$ thin films as a function of the IrMn$_3$ film thickness.

**Figure S4**. AMR effect for the MgO(001)/Ni$_3$Fe(2.5 nm)/IrMn$_3$ (5.2 nm)/Ni$_3$Fe(15 nm)/CoO(2.0 nm)/Pt(3.5 nm) heterostructure obtained at 300 K.

**Figure S5**. Magneto-optical detection of multiple MFC states.

**Figure S6.** Temperature-dependent MOKE hysteresis loops obtained in various MFC states from the front side of the sample.

**Figure S7.** Magnetic and magnetoresistance properties of MgO(001)/Ni$_3$Fe(2.5 nm)/IrMn$_3$(5.2 nm)/Ni$_3$Fe(15 nm)/Pt(3.5 nm)  heterostructures in various MFC states.

**Figure S8.** Temperature-dependent transition of the EB field in MgO(001)/Ni$_3$Fe(2.5 nm)/IrMn$_3$(5.2 nm)/Ni$_3$Fe(15 nm)/Pt(3.5 nm) heterostructure.

**Figure S9.** Magnetic and magnetotransport properties of MgO(001)/Ni$_3$Fe(2.5 nm)/IrMn$_3$(16 nm)/Ni$_3$Fe(15 nm)/CoO(2.0 nm)/Pt(3.5 nm) heterostructures in various MFC states.

**Figure S10.** Magnetic properties of MgO(001)/Ni$_3$Fe(2.5 nm)/ IrMn$_3$(16 nm)/Ni$_3$Fe(15 nm)/Pt(3.5 nm) heterostructures in various MFC states.

**Figure S11.** Coercive and exchange bias field as a function of antiferromagnetic ($\gamma$-IrMn$_3$) film thickness in MgO(001)/IrMn$_3$/ Ni$_3$Fe bilayers at 300 K.

**Figure S12.** Collinear and spin-flop type magnetization reversal behavior of MgO(001)/IrMn$_3$/Ni$_3$Fe bilayers as a function of antiferromagnetic ($\gamma$-IrMn$_3$) film thickness.

**Figure S13.** Degeneracy of Exchange bias field as a function of antiferromagnetic ($\gamma$-IrMn$_3$) film thickness in MgO(001)/IrMn$_3$/ Ni$_3$Fe bilayers at 80 K.

**Figure S14.** Experimental geometry and the sample structure for XAS/ XMCD experiments.

**Figure S15.** XAS/XMCD spectra of Co, Ni, and Fe.

**Figure S16.** Element-specific Magnetic hysteresis loops obtained near the resonant Co and Ni $L_3$-edges at 300 K.

**Figure S17.** Element-specific magnetic hysteresis loops at 80 K in parallel MFC state ($\mu_0 H_{FC}$ = +100 mT).

**Figure S18.** Element-specific magnetic hysteresis loops obtained at 80 K in an anti-parallel MFC state ($\mu_0 H_{FC}$ = -100 mT).

**Figure S19.** Magnetic field stability of Antiferromagnetic Mn spin structure at 80 K in parallel and anti-parallel MFC states.

**Table S1.** Summary of various interfacial properties of the exchange coupled heterostructures measured at 300 K and 80 K, extracted from the element-specific XMCD hysteresis loops shown in Supplementary Figures S16-S19.

**Author Contributions**

MA and SSPKA proposed the experiments. SSPKA prepared the samples with supervision from MA and OH. FG performed the XRD scans, and SSPKA performed the analysis. FS performed the Atomic force microscopy scans, and SSPKA analyzed the images. SSPKA and DB performed the magnetic and transport measurements. SSPKA performed the MOKE hysteresis loops and SKerrM imaging. SSPKA performed the element-specific XAS/XMCD measurements with assistance from CL. SSPKA wrote the manuscript with comments from all the co-authors.

**Notes**

The authors declare no competing financial interests.

**ACKNOWLEDGMENT**

The authors thank Dr. Birgit Hebler from the University of Augsburg, Dr. Florin Radu and the staff at Helmholtz-Zentrum Berlin on beamline PM2 VEKMAG end station for providing technical support during the synchrotron beamtime. The authors thank Manuel Monecke from TU Chemnitz for performing XPS measurements. Additionally, the authors acknowledge Dr. Martin Dehnert and Prof. Robert Magerle from TU Chemnitz for facilitating access to the atomic force microscopy lab. The authors are thankful to Dr. Jürgen Lindner from HZDR, Dr. Rodolfo Gallardo from CEDENNA, and Dr. Patrick Matthes from Fraunhofer ENAS for valuable discussions. GS, DB, and DRTZ acknowledge the Deutsche Forschungsgemeinschaft for the funding (project FOR 1154).

## REFERENCES

(1) Železný, J.; Wadley, P.; Olejník, K.; *A.* Hoffmann; H. Ohno. Spin transport and spin torque in antiferromagnetic devices. *Nature Phys* 14**,** 220–228 (**2018**).

(2) Wadley, P.; Howells, B.; Železný, J.; Andrews, C.; Hills, V.; Campion, R. P.; Novák, V.; Olejník, K.; Maccherozzi, F.; Dhesi, S. S.; Martin, S. Y.; Wagner, T.; Wunderlich, J.; Freimuth, F.; Mokrousov, Y.; Kuneš, J.; Chauhan, J. S; Grzybowski, M. J.; Rushforth, A. W.; Edmonds, K. W.; Gallagher, B. L.; Jungwirth. T. Electrical switching of an antiferromagnet. Science 351, 587–590 (**2016**).

(3) S. Yu. Bodnar, Y. Skourski, O. Gomonay, J. Sinova, M. Kläui, and M. Jourdan. Magnetoresistance Effects in the Metallic Antiferromagnet $Mn_2Au$. Phys. Rev. Applied 14, 014004, (**2020**).

(4) Olejník, K.; Seifert, T.; Kašpar, Z.; Novák, V.; Wadley, P.; Campion, R. P.; Baumgartner, M.; Gambardella, P.; Němec, P.; Wunderlich, J.; Sinova, J.; Kužel, P.; Müller, M.; Kampfrath, T.; Jungwirth, T. Terahertz electrical writing speed in an antiferromagnetic memory. Science Advances 4, eaar3566, (**2018**).

(5) Shi, J.; Lopez-Dominguez, V.; Garesci, F.; Wang, C.; Almasi, H.; Grayson, M.; Finocchio, G.; Amiri, K. P. Electrical manipulation of the magnetic order in antiferromagnetic PtMn pillars. Nat Electron 3, 92–98 (**2020**).

(6) Galkina, E. G.; Ivanov, B. A. Dynamic solitons in antiferromagnets (Review Article). Low Temperature Physics 44, 618 (**2018**).

(7) Bodnar, Yu. S.; Šmejkal, L.; Turek, I.; Jungwirth, T.; Gomonay, O.; Sinova, J.; Sapozhnik. A. A.; Elmers, H.-J.; Kläui, M.; Jourdan, M. Writing and reading antiferromagnetic $Mn_2Au$ by Néel spin-orbit torques and large anisotropic magnetoresistance. Nat Commun 9, 348 (**2018**).

(8) Liu, Z. Q.; Chen, H.; Wang, J. M.; Liu, J. H.; Wang, K.; Feng, Z. X.; Yan, H.; Wang, X. R.; Jiang, C. B.; Coey, J. M. D.; MacDonald, A. H. Electrical switching of the topological anomalous Hall effect in a non-collinear antiferromagnet above room temperature. Nature Electronics 1, 172–177 (**2018**).

(9) Takeuchi, Y.; Yamane, Y.; Yoon, J.-Y.; Itoh, R.; Jinnai, B.; Kanai, S.; Ieda, J.; Fukami, S.; Ohno, H. Chiral-spin rotation of non-collinear antiferromagnet by spin–orbit torque. Nat. Mater. (**2021**). https://doi.org/10.1038/s41563-021-01005-3

(10) Tsai, H.; Higo, T.; Kondou, K.; Nomoto, T.; Sakai, A.; Kobayashi, A.; Nakano, T.; Yakushiji, K.; Arita, R.; Miwa, S.; Otani, Y.; Nakatsuji, S. Electrical manipulation of a topological antiferromagnetic state. Nature 580, 608–613 (**2020**).

(11) Arpaci, S.; Lopez-Dominguez, V.; Shi, J.; Sánchez-Tejerina, L.; Garesci, F.; Wang, C.; Yan, X.; Sangwan, V. K.; Grayson, M. A.; Hersam, M. C.; Finocchio, G.; Amiri, P. K. Observation of current-induced switching in non-collinear antiferromagnetic $IrMn_3$ by differential voltage measurements. Nat Commun 12, 3828 (**2021**).

(12) Moriyama, T.; Oda, K.; Ohkochi, T.; Kimata, M.; Ono, T. Spin torque control of antiferromagnetic moments in NiO. Sci. Rep. 8, 14167 (**2018**).

(13) Baldrati, L.; Schmitt, C.; Gomonay, O.; Lebrun, R.; Ramos, R.; Saitoh, E.; Sinova, J.; Kläui, M. Efficient Spin Torques in Antiferromagnetic CoO/Pt Quantified by Comparing Field- and Current-Induced Switching. Phys. Rev. Lett. 125, 077201 (**2020**)

(14) Železný, J.; Gao, H.; Výborný, K.; Zemen, J.; Mašek, J.; Manchon, A.; Wunderlich, J.; Sinova, J.; Jungwirth, T. Relativistic Néel-Order Fields Induced by Electrical Current in Antiferromagnets. Phys. Rev. Lett. 113, 157201 (**2014**).

(15) Chiang, C. C.; Huang, S. Y.; Qu, D.; Wu, P. H.; Chien, C. L. Absence of evidence of electrical switching of the antiferromagnetic Néel vector. Phys Rev. Lett. 123, 227203 (**2019**).

(16) Churikova, A.; Bono, D.; Neltner, B.; Wittmann, A.; Scipioni, L.; Shepard, A.; Newhouse-Illige, T.; Greer, J.; Beach. G. S. D. Non-magnetic origin of spin Hall magnetoresistance-like signals in Pt films and epitaxial NiO/Pt bilayers featured. Appl. Phys. Lett. 116, 022410 (**2020**).

(17) Cogulu, E.; Statuto, N. N.; Cheng, Y.; Yang, F.; Chopdekar, R. V.; Ohldag, H.; Kent, A. D. Direct imaging of electrical switching of antiferromagnetic Néel order in α−Fe2O3 epitaxial films. Phys. Rev. B 103, L100405 (**2021**).

(18) Reimers, S.; Grzybowski, M. J.; Andrews, C.; Wang, M.; Chauhan, J. S.; Gallagher, B. L.; Campion, R. P.; Edmonds, K. W.; Dhesi, S. S.; Maccherozzi, F.; Novak, V.; Wunderlich, J.; Jungwirth, T. Current polarity-dependent manipulation of antiferromagnetic domains. Nat. Nanotech. 13, 362 (**2018**).

(19) Zink, B. The Heat in Antiferromagnetic Switching. Physics 12, 134 (**2019**).

(20) Meer, H.; Schreiber, F.; Schmitt, C.; Ramos, R.; Saitoh, E.; Gomonay, O.; Sinova, J.; Baldrati, L.; Kläui, M. Direct Imaging of Current-Induced Antiferromagnetic Switching Revealing a Pure Thermomagnetoelastic Switching Mechanism in NiO. Nano Lett. 21, 1, 114–119 (**2021**).

(21) Marti, X.; Fina, I.; Frontera, C.; Liu, J.; Wadley, P.; He, Q.; Paull, R. J.; Clarkson, J. D.; Kudrnovský, J.; Turek, I.; Kuneš, J.; Yi, D.; Chu, J-H.; Nelson, C. T.; You, L.; Arenholz, E.; Salahuddin, S.; Fontcuberta, J.; Jungwirth, T.; Ramesh, R. Room-temperature antiferromagnetic memory resistor. Nature Materials volume 13, 367–374 (**2014**).

(22) Song, C.; You, Y.; Chen, X.; Zhou, X.; Wang, Y.; Pan, F. How to manipulate magnetic states of antiferromagnets, *Nanotechnology* 29 112001 (**2018**).

(23) Park, B. G.; Wunderlich, J.; Martí, X.; Holý, V.; Kurosaki, Y.; Yamada, M.; Yamamoto, H.; Nishide, A.; Hayakawa, J.; Takahashi, H.; Shick, A. B.; Jungwirth. T. A spin-valve-like magnetoresistance of an antiferromagnet-based tunnel junction. Nature Materials 10, 347–351 (**2011**).

(24) Nogués, J.; Schuller, I. K. Exchange bias. J. Magn. Magn. Mater. 192, 203–32 (**1999**).

(25) Radu, F.; Zabel, H. Exchange bias effect of ferro- /antiferromagnetic heterostructure, Springer Tr. Mod. Phys. 227, 97 (**2008**).

(26) O'Grady,' K.; Fernandez-Outon, L. E.; Vallejo-Fernandez. G. A new paradigm for exchange bias in polycrystalline thin films. J. Magn. Magn. Mater. 322, 883-899 (**2010**).

(27) Ali, M.; Marrows, C. H.; Hickey, B. J. Onset of exchange bias in ultrathin antiferromagnetic layers. Phys. Rev. B 67, 172405 (**2003**).

(28) Ali, M.; Marrows, C. H. ; Al-Jawad, M.; Hickey, B. J.; Misra, A.; Nowak, U.; Usadel, K. D. Antiferromagnetic layer thickness dependence of the IrMn/Co exchange-bias system. Phys. Rev. B 68, 214420 (2003).

(29) Brück, S.; Sort, J.; Baltz, V.; Suriñach, S.; Muñoz, J.; Dieny, B.; Baró, M.; Nogués, J. Exploiting Length Scales of Exchange-Bias Systems to Fully Tailor Double-Shifted Hysteresis Loops. Adv. Mater., 17: 2978-2983 (2005)

(30) Martí, X.; Park, B. G.; Wunderlich, J.; Reichlová, H.; KuroSaki, Y.; Yamada, M.; Yamamoto, H.; Nishide, A.; Hayakawa, J.; Takahashi, H.; Jungwirth, T. Electrical Measurement of Antiferromagnetic Moments in Exchange-Coupled IrMn/NiFe Stacks. Phys. Rev. Lett. 108, 017201(**2012**).

(31) Jenkins, S.; Fan, W. J.; Gaina, R.; Chantrell, R. W.; Klemmer, T.; Evans; R. F. L. Atomistic origin of exchange anisotropy in noncollinear $\gamma-$IrMn$_3$–CoFe bilayers. Phys. Rev. B 102, 140404 (**2020**).

(32) Ohldag, H.; Scholl, A.; Nolting, F.; Arenholz, E.; Maat, S.; Young, A. T.; Carey, M.; Stöhr, J. Correlation Between Exchange Bias and Pinned Interfacial Spins. Phys. Rev. Lett. 91, 017203 (**2003**).

(33) Camarero, J.; Sort, J.; Hoffmann, A.; García-Martín, J. M.; Dieny, B.; Miranda, R.; Nogués, J. Origin of the Asymmetric Magnetization Reversal Behavior in Exchange-Biased Systems: Competing Anisotropies. Phys. Rev. Lett. 95, 057204 (**2005**).

(34) Brems, S.; Temst, K.; Haesendonck, C. V. Origin of the Training Effect and Asymmetry of the Magnetization in Polycrystalline Exchange Bias Systems, Phys. Rev. Lett. 99, 067201 (**2007**).

(35) Leighton, C.; Nogués, J.; Jönsson-Åkerman, B. J.; Schuller, I. K. Coercivity Enhancement in Exchange Biased Systems Driven by Interfacial Magnetic Frustration. Phys. Rev. Lett. 84, 3466 (**2000**).

(36) Jenkins, S.; Chantrell, R. W.; Evans, R. F. L. Atomistic origin of the athermal training effect in granular IrMn/CoFe bilayers. Phys. Rev. B 103, 104419 (2021).

(37) Guo, J.; Zhao, X.; Lu, Z.; Shi, P.; Tian, Y.; Chen, Y.; Yan, S.; Bai, L.; Harder, M. Evidence for linear dependence of exchange bias on pinned uncompensated spins in an Fe/FeO bilayer. Phys. Rev. B 103, 054413 (**2021**).

(38) Ali, M.; Adie, P.; Marrows, C.; Greig D.; Hickey, B. J.; Stamps R. L; Exchange bias using a spin glass. *Nature Mater* 6, 70–75 (**2007**).

(39) E. Maniv; R. A. Murphy; S. C. Haley; S. Doyle; C. John; A. Maniv; S. K. Ramakrishna; Y.-L. Tang; P. Ercius; R. Ramesh; A. P. Reyes; J. R. Long; Analytis, J. G. Exchange bias due to coupling between coexisting antiferromagnetic and spin-glass orders. *Nat. Phys.* **17,** 525–530 (**2021**).

(40) Binder, K.; Young, A. P.; Spin glasses: Experimental facts, theoretical concepts, and open questions. Rev. Mod. Phys. 58, 801 (**1986**).

(41) Ma, X.; Fang, F.; Li, Q.; Zhu, J.; Yang, Y.; Wu, Y. Z.; Zhao, H. B.; Lüpke, G. Ultrafast spin exchange-coupling torque via photo-excited charge-transfer processes. Nat Commun 6, 8800 (**2015**).

(42) Ma, X.; Yu, G.; Razavi, S. A.; Sasaki, S. S.; Li, X.; Hao, K.; Tolbert, S. H.; Wang, K. L.; Li, X. Dzyaloshinskii-Moriya Interaction across an Antiferromagnet-Ferromagnet Interface. Phys. Rev. Lett. 119, 027202 (**2017**).

(43) Kurenkov, A.; Gupta, S. D.; Zhang, C.; Fukami, S.; Horio, Y. ; Ohno, H. Artificial Neuron and Synapse Realized in an Antiferromagnet/Ferromagnet Heterostructure Using Dynamics of Spin–Orbit Torque Switching. Adv. Mater. 31, 1900636 (2019).

(44) Vallobra, P.; Fache, T.; Xu, Y.; Zhang, L.; Malinowski, G.; Hehn, M.; Rojas-Sánchez, J.-C.; Fullerton, E. E.; Mangin, S. Manipulating exchange bias using all-optical helicity-dependent switching. Phys. Rev. B 96, 144403 (**2017**).

(45) Liao, Yu-Ch.; Nikonov, D. E.; Dutta, S.; Chang, S.-C.; Hsu, C.-S.; Young, I. A.; Naeemi, A. Understanding the Switching Mechanisms of the Antiferromagnet/Ferromagnet Heterojunction. Nano Letters 20 (11), 7919-7926 (**2020**).

(46) Zhou, J.; Shu, X.; Liu, Y.; Wang, X.; Lin, W.; Chen, S.; Liu, L.; Xie, Q; Hong, T.; Yang, P.; Yan, B.; Han, X.; Chen, J. Magnetic asymmetry induced anomalous spin-orbit torque in IrMn. Phys. Rev. B 101, 184403 (**2020**).

(47) Szunyogh, L.; Lazarovits, B. ; Udvardi, L.; Jackson, J.; Nowak, U. Giant magnetic anisotropy of the bulk antiferromagnets IrMn and IrMn$_3$ from first principles. Phys. Rev. B 79, 020403 (**2009**).

(48) Yamaoka, T.; Mekata, M.; Takaki, H. Neutron diffraction study of γ-phase Mn-Ir single crystals. J. Phys. Soc. Japan, 36, 438–444 (1974).

(49) Sakuma, A.; Fukamichi, K.; Sasao, K.; Umetsu. R. Y. First-principles study of the magnetic structures of ordered and disordered Mn-Ir alloys. Phys. Rev. B 67, 024420 (**2003**).

(50) Shick, A. B.; Khmelevskyi, S.; Mryasov, O. N.; Wunderlich, J.; Jungwirth, T. Spin-orbit coupling induced anisotropy effects in bimetallic antiferromagnets: A route towards antiferromagnetic spintronics. Phys. Rev. B 81, 212409 (**2010**).

(51) Kohn, A.; Kovács, A.; Fan, R.; McIntyre, G. J.; Ward, R. C. C.; Goff, J. P. The antiferromagnetic structures of IrMn$_3$ and their influence on exchange-bias. *Sci Rep* **3,** 2412 (**2013**).

(52) Vallejo-Fernandez, G.; Aley, N. P.; Fernandez-Outon, L. E.; O'Grady, K. Control of the setting process in CoFe/IrMn exchange bias systems. J. Appl. Phys. 104, 033906 (**2008**).

(53) Jenkins, S; Chantrell, R. W.; Evans, R. F. L. Exchange bias in multigranular noncollinear IrMn$_3$/CoFe thin films. Phys. Rev. B 103, 014424 (**2021**).

(54) Tshitoyan, V.; Ciccarelli, C.; Mihai, A. P.; Ali, M.; Irvine, A. C.; Moore, T. A.; Jungwirth, T.; Ferguson, A. J. Electrical manipulation of ferromagnetic NiFe by antiferromagnetic IrMn. Phys. Rev. B 92, 214406 (**2015**).

(55) Arekapudi, S. S. P. K.; Böhm, B.; Ramasubramanian, L.; Ganss, F.; Heinig, P; Stienen, S.; Fowley, C.; Lenz, K.; Deac, A. M.; Albrecht, M.; Hellwig, O. Direct imaging of distorted vortex structures and magnetic vortex annihilation processes in ferromagnetic/antiferromagnetic disk structures. Phys. Rev. B 103, 014405 (**2021**).

(56) Gupta, K.; Wesselink, R. J. H.; Liu, R.; Yuan, Z.; Kelly, P. J. Disorder Dependence of Interface Spin Memory Loss. Phys. Rev. Lett. 124, 087702 (**2020**).

(57) Perzanowski, M.; Marszalek, M,; Zarzycki, A.; Krupinski, M.; Dziedzic, A.; Zabila, Y. Influence of Superparamagnetism on Exchange Anisotropy at CoO/[Co/Pd] Interfaces. ACS Applied Materials & Interfaces 8 (41), 28159-28165 (**2016**).

(58) Cramer, J., Fuhrmann, F., Ritzmann, U;  Gall, V.; Niizeki, T.; Ramos, R.; Qiu, Z.; Hou, D.; Kikkawa, T.; Sinova, J.; Nowak, U.; Saitoh, E.; Kläui, M. Magnon detection using a ferroic collinear multilayer spin valve. *Nat Commun* 9, 1089 (**2018**).

(59) Qiu, Z.; Hou, D.; Barker, J.; Yamamoto, K.; Gomonay, O.; Saitoh, E. Spin colossal magnetoresistance in an antiferromagnetic insulator. Nature Mater 17, 577–580 (**2018**).

(60) Fernandez-Outon, L. E.; Vallejo-Fernandez, G.; Manzoor, S.; Hillebrands, B.; O'Grady, K. Interfacial spin order in exchange biased systems. J. Appl. Phys. 104, 093907 (**2008**).

(61) Appel, P.; Shields, B. J.; Kosub, T.; Hedrich, N.; Hübner, R.; Faßbender, J; Makarov, D.; Maletinsky, P. Nanomagnetism of Magnetoelectric Granular Thin-Film Antiferromagnets. Nano Lett. 19, 3, 1682–1687 (**2019**).

(62) Taylor, J. M.; Lesne, E.; Markou, A.; Dejene, F. K; Sivakumar, P. K.; Pöllath, S.; Rana, K. G.; Kumar, N.; Luo, C.; Ryll, H.; Radu, F.; Kronast, F.; Werner, P.; Back, C. H.; Felser, C.; Parkin, S. S. P. Magnetic and electrical transport signatures of uncompensated moments in epitaxial thin films of the noncollinear antiferromagnet $Mn_3Ir$. Appl. Phys. Lett. 115, 062403 (**2019**).

(63) Binek, Ch.; Hochstrat, A.; Kleemann, W. Exchange bias in a generalized Meiklejohn–Bean approach J. Magn. Magn. Mater. 234, 353 (**2001**).

(64) Kobrinskii, A. L.; Goldman, A. M; Varela, M.; Pennycook, S. J.; Thickness dependence of the exchange bias in epitaxial manganite bilayers. Phys. Rev. B 79, 094405 (**2009**).

(65) Wang, Y. Y.; Song, C.; Cui, B.; Wang, G. Y.; Zeng, F.; Pan, F. Room-Temperature Perpendicular Exchange Coupling and Tunneling Anisotropic Magnetoresistance in an Antiferromagnet-Based Tunnel Junction. Phys. Rev. Lett. 109, 137201 (**2012**).

(66) Malozemoff, A. P. Random-field model of exchange anisotropy at rough ferromagnetic-antiferromagnetic interfaces. Phys. Rev. B 35, 3679 (**1987**).

(67) Mauri, D.; Siegmann, H. C.; Bagus, P. S.; Kay, E. Simple model for thin ferromagnetic films exchange coupled to an antiferromagnetic substrate. J. Appl. Phys. 62, 3047 (**1987**).

(68) Miccoli, I.; Edler, F.; Pfnür, H.; Tegenkamp, C. The 100th anniversary of the four-point probe technique: the role of probe geometries in isotropic and anisotropic systems. J. Phys.: Condens. Matter 27 (**2015**) 223201 (29pp).

(69) Tatsumoto, E.; Kuwahara, K.; Goto, M. Magnetoresistance Effect in the Magnetization Reversal of Permalloy Films. J. Phys. Soc. Jpn. 15, pp. 1703-1703 (**1960**).

(70) Jauch, W.; Reehuis, M.; Bleif, H. J.; Kubanek, F.; Pattison, P. Crystallographic symmetry and magnetic structure of CoO. Phys. Rev. B 64, 052102 (**2001**).

(71) Anderson, P. W. Antiferromagnetism. Theory of Superexchange Interaction. Phys. Rev. 79, 350 (**1950**).

(72) Lange, F.; Martin, M. The electrical conductivity of CoO: Experimental results and a new conductivity model. Berichte der Bunsengesellschaft für physikalische Chemie, 101: 176-184 (**1997**).

(73) O. Yıldırım; D. Hilliard; S. S. P. K. Arekapudi; C. Fowley; H. Cansever; L. Koch; L. Ramasubramanian; S. Zhou; R. Böttger; J. Lindner; J. Faßbender; O. Hellwig; A. M. Deac. Ion-Irradiation-Induced Cobalt/Cobalt Oxide Heterostructures: Printing 3D Interfaces. *ACS Appl. Mater. Interfaces* 2020, 12, 8, 9858–9864 (**2020**).

(74) Berger, L. Low-field magnetoresistance and domain drag in ferromagnets. J. Appl. Phys. 49, 2156 (**1978**).

(75) Viret, M.; Vignoles, D.; Cole, D.; Coey, J. M. D.; Allen, W.; Daniel, D. S; Gregg, J. F. Spin scattering in ferromagnetic thin films. Phys. Rev. B 53, 8464 (**1996**).

(76) Levy, P. M.; Zhang, S. Resistivity due to Domain Wall Scattering. Phys. Rev. Lett. 79, 5110 (**1997**).

(77) Tatara, G; Fukuyama, H. Resistivity due to a Domain Wall in Ferromagnetic Metal. Phys. Rev. Lett. 78, 3773 (**1997**).

(78) McGuire, T. R.; Potter, R. I. Anisotropic magnetoresistance in ferromagnetic 3d alloys. IEEE Trans. Magn. 11, 1018–1038 (**1975**).

(79) Schuller, I. K.; Morales, R.; Batlle, X.; Nowak, U.; Güntherodte, G. Role of the antiferromagnetic bulk spins in exchange bias. J. Magn. Magn. Mater 416, 2-9 (**2016**).

(80) Guo, Z. B.; Wu, Y. H.; Qiu, J. J.; Zong, B. Y.; Han, G. C. Exchange bias and magnetotransport properties in IrMn/NiFe/FeMn structures. Phys. Rev. B 78, 184413 (**2008**).

# SUPPLEMENTARY MATERIAL

# CONTENTS

**Figure S1**. Magnetization of the FM NiFe(15 nm) layer as a function of temperature for the $\gamma$-IrMn$_3$/Ni$_3$Fe bilayers.

**Figure S2**. Evolution of the Ni$_3$Fe(15 nm) domain structure as a function of the IrMn$_3$ film thickness.

**Figure S3**. Tetragonality ($c/a$-ratio) of the Ni$_3$Fe and IrMn$_3$ thin films as a function of the IrMn$_3$ film thickness.

**Figure S4**. AMR effect for the MgO(001)/Ni$_3$Fe(2.5 nm)/IrMn$_3$ (5.2 nm)/Ni$_3$Fe(15 nm)/CoO(2.0 nm)/Pt(3.5 nm) heterostructure obtained at 300 K.

**Figure S5**. Magneto-optical detection of multiple MFC states.

**Figure S6**. Temperature-dependent MOKE hysteresis loops obtained in various MFC states from the front side of the sample.

**Figure S7**. Magnetic and magnetoresistance properties of MgO(001)/Ni$_3$Fe(2.5 nm)/IrMn$_3$(5.2 nm)/Ni$_3$Fe(15 nm)/Pt(3.5 nm) heterostructures in various MFC states.

**Figure S8**. Temperature-dependent transition of the EB field in MgO(001)/Ni$_3$Fe(2.5 nm)/IrMn$_3$(5.2 nm)/Ni$_3$Fe(15 nm)/Pt(3.5 nm) heterostructure.

**Figure S9**. Magnetic and magnetotransport properties of MgO(001)/Ni$_3$Fe(2.5 nm)/IrMn$_3$(16 nm)/Ni$_3$Fe(15 nm)/CoO(2.0 nm)/Pt(3.5 nm) heterostructures in various MFC states.

**Figure S10**. Magnetic properties of MgO(001)/Ni$_3$Fe(2.5 nm)/ IrMn$_3$(16 nm)/Ni$_3$Fe(15 nm)/Pt(3.5 nm) heterostructures in various MFC states.

**Figure S11**. Coercive and exchange bias field as a function of antiferromagnetic ($\gamma$-IrMn$_3$) film thickness in MgO(001)/IrMn$_3$/ Ni$_3$Fe bilayers at 300 K.

**Figure S12**. Collinear and spin-flop type magnetization reversal behavior of MgO(001)/IrMn$_3$/Ni$_3$Fe bilayers as a function of antiferromagnetic ($\gamma$-IrMn$_3$) film thickness.

**Figure S13**. Degeneracy of Exchange bias field as a function of antiferromagnetic ($\gamma$-IrMn$_3$) film thickness in MgO(001)/IrMn$_3$/ Ni$_3$Fe bilayers at 80 K.

**Figure S14**. Experimental geometry and the sample structure for XAS/ XMCD experiments.

**Figure S15**. XAS/XMCD spectra of Co, Ni, and Fe.

**Figure S16**. Element-specific Magnetic hysteresis loops obtained near the resonant Co and Ni $L_3$-edges at 300 K.

**Figure S17**. Element-specific magnetic hysteresis loops at 80 K in parallel MFC state ($\mu_0 H_{FC}$ = +100 mT).

**Figure S18**. Element-specific magnetic hysteresis loops obtained at 80 K in an anti-parallel MFC state ($\mu_0 H_{FC}$ = -100 mT).

**Figure S19**. Magnetic field stability of Antiferromagnetic Mn spin structure at 80 K in parallel and anti-parallel MFC states.

**Table S1.** Summary of various interfacial properties of the exchange coupled heterostructures measured at 300 K and 80 K, extracted from the element-specific XMCD hysteresis loops shown in Supplementary Figures S16-S19.

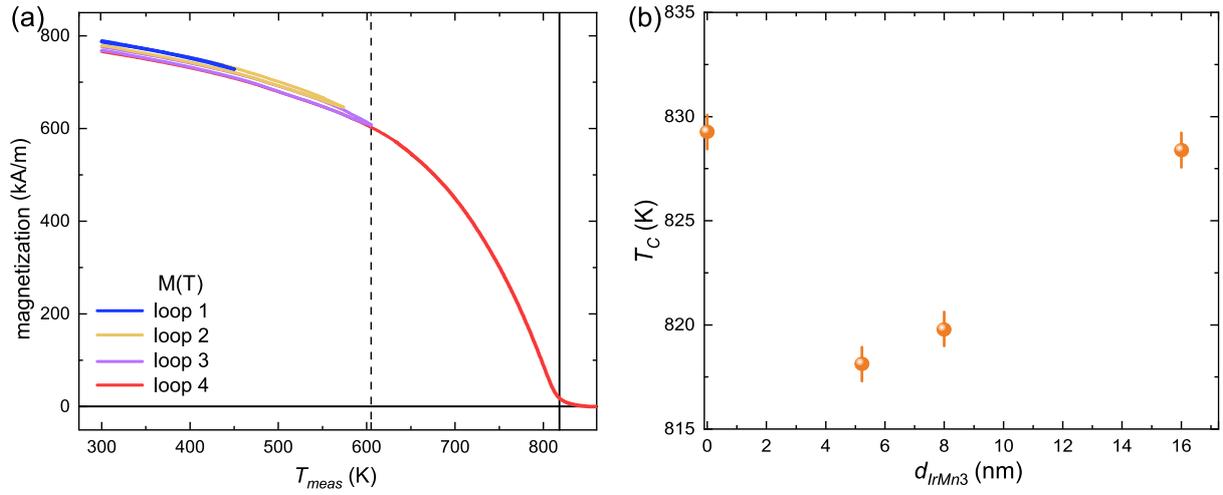

**Figure S1. Magnetization of the FM NiFe(15 nm) layer as a function of temperature for the γ-IrMn₃/Ni₃Fe bilayers**. **(a)** $M$(T) curves for the MgO(001)/IrMn₃(5.2 nm)/Ni₃Fe (15 nm)/Al₂O₃ (2.0 nm) sample. Minor $M$(T) loops indicate a subtle decrease in the magnetization ($\Delta M \approx 32$ kA/m at 300 K) of the Ni₃Fe layer as a consequence of heating the sample to 450 K (loop 1), 575 K (loop 2), 605 K (loop 3), and 865 K (loop 4), respectively. Interdiffusion at the IrMn₃/Ni₃Fe interface can result in the loss of magnetization.[1] **(b)** Curie temperature $T_C$ as a function of IrMn₃ thickness $d_{IrMn3}$ in MgO(001)/IrMn₃('$d_{IrMn3}$' nm)/Ni₃Fe (15 nm) bilayers. At $d_{IrMn3}$ = 5.2 nm, $T_C$ tends to drop by ~11 K, most likely due to the active interdiffusion at the IrMn₃/Ni₃Fe interface, resulting from heating the bilayer sample to elevated temperature (>820 K).

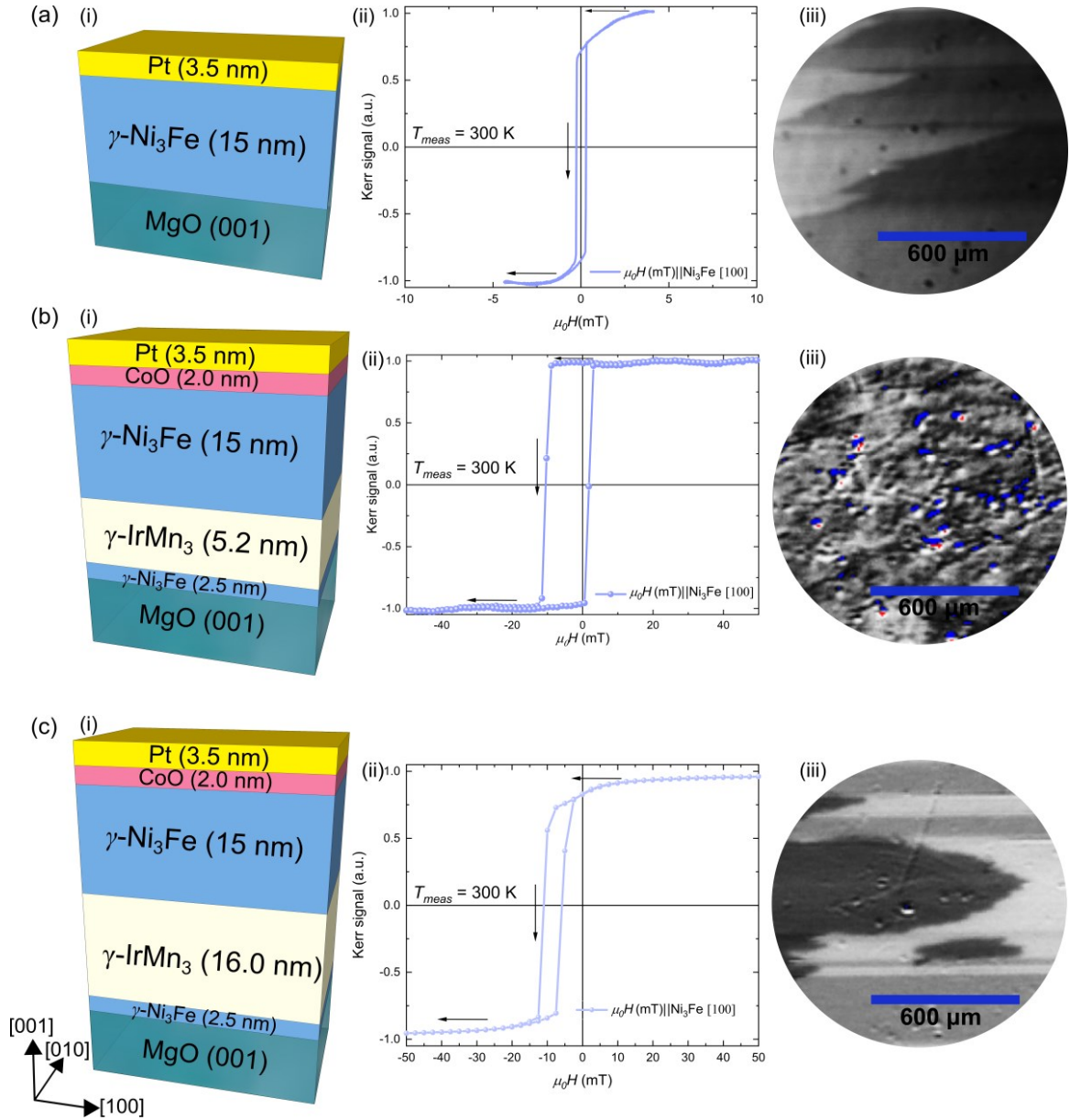

**Figure S2. Evolution of the Ni₃Fe(15 nm) domain structure as a function of the IrMn₃ film thickness (a)(i)**
Reference NiFe(15 nm) sample, **(a)(ii)** Longitudinal MOKE hysteresis loop obtained at 300 K from the front side
of the sample, and **(a)(iii)** magnetic domain structure of the sample near the coercive field $H_C$ obtained using
SKerrM imaging in longitudinal geometry. **(b)(i)** Exchange coupled heterostructure with $d_{IrMn}$ = 5.2 nm, and
**(b)(iii)** magnetic domain structure of the sample near $H_C$ indicates a fractal domain structure. **(c)(i)** Exchange
coupled heterostructure with $d_{IrMn}$ = 16.0 nm, **(c)(ii)** MOKE hysteresis loop obtained at 300 K, and **(c)(iii)**
corresponding magnetic domain structure of the sample obtained near $H_C$. Due to the limited penetration depth of
the MOKE laser, the MOKE hysteresis loops and the magnetic domain structure observed here correspond to the top FM Ni$_3$Fe(15 nm) layer. In multilayers with a thinner IrMn layer ($d_{IrMn}$ = 5.2 nm), the ferromagnetic Ni$_3$Fe(15 nm) thin film exhibits (discontinuous) fractal magnetic domains accompanied by an increase in $H_C$. This is most likely due to the presence of reversible and frustrated AFM-IrMn grains as a consequence of tetragonality (*c/a*-ratio), atomic site disorder, and lower grain volume of the IrMn$_3$(5.2 nm) thin film. However, increasing the IrMn film thickness results again in larger magnetic domains with somewhat continuous domain propagation and a reduction in $H_C$, most likely due to the tetragonal strain relaxation in the heterostructures, as shown below.

**Note**: For the reference MgO(100)/NiFe(15 nm) thin films, we calculated the upper limit of the domain wall width $\delta_N = \pi\sqrt{A_{ex}/K_u}$ = 121.16 nm, where $A_{ex}$ = 1.3 x 10$^{-12}$ J/m, and $K_u$ = 0.874 kJ/m$^3$. In the NiFe films with thickness <30-35 nm, Néel-type domain walls are expected.[2]

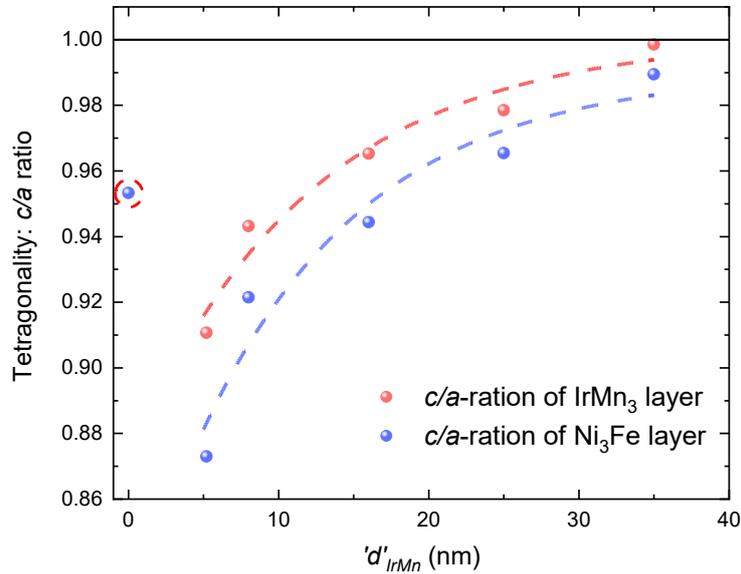

**Figure S3. Tetragonality (*c/a*-ratio) of the Ni$_3$Fe and IrMn$_3$ thin films as a function of the IrMn$_3$ film thickness.** The in-plane (*a*) and out-of-plane (*c*) lattice parameters of the Ni$_3$Fe and IrMn$_3$ thin films were obtained from the Bragg scans. The tetragonality of Ni$_3$Fe(15 nm) thin films with $d_{IrMn}$ = 0 and 16 nm are quite similar; this might explain the reason for the observation of somewhat similar magnetic domain structure in both samples shown in (a)(iii) and (c)(iii), respectively.

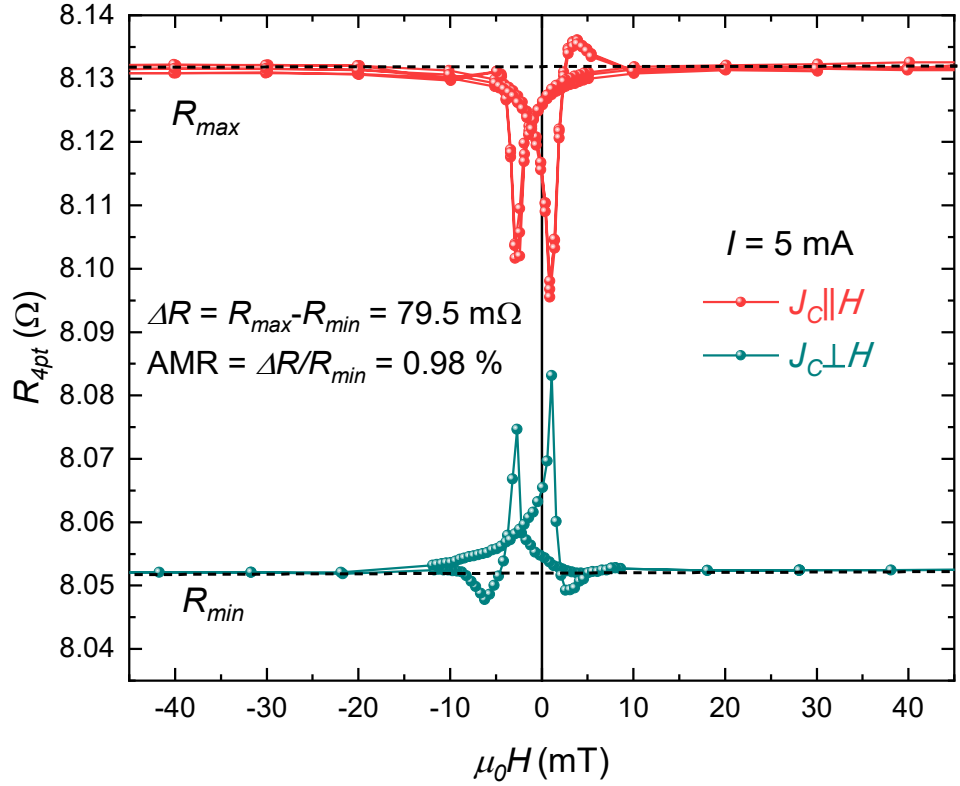

**Figure S4. AMR effect for MgO(001)/Ni₃Fe(2.5 nm)/IrMn₃(5.2 nm)/Ni₃Fe(15 nm)/CoO(2.0 nm)/Pt(3.5 nm) heterostructure obtained at 300 K.** Magnetoresistance curves obtained at $I$ = 5 mA ($J_C \| \mu_0 H \|$[100] (in red) and $J_C \|$[010]$\perp \mu_0 H \|$[100] (in green)) for the sample shown in Figure 6(a) in the main paper. The AMR = 0.98% is significantly less than the typical bulk values (4-5%) reported for Ni₈₀Fe₂₀ alloys,[3] most likely due to the presence of extrinsic contributions, namely grain boundary scattering, interfacial scattering, and strain-induced effects. Furthermore, we found an increase in resistivity $\rho$ = 23.3 μΩ·cm of the Ni₃Fe(15 nm) thin films deposited on the IrMn(5.2 nm) layer and $\rho$ = 21.6 μΩ·cm for Ni₃Fe(15 nm) thin films deposited directly on MgO(001) single crystal substrate, compared to the reference bulk Ni₈₀Fe₂₀ $\rho$ ≈ 14 μΩ·cm.[3]

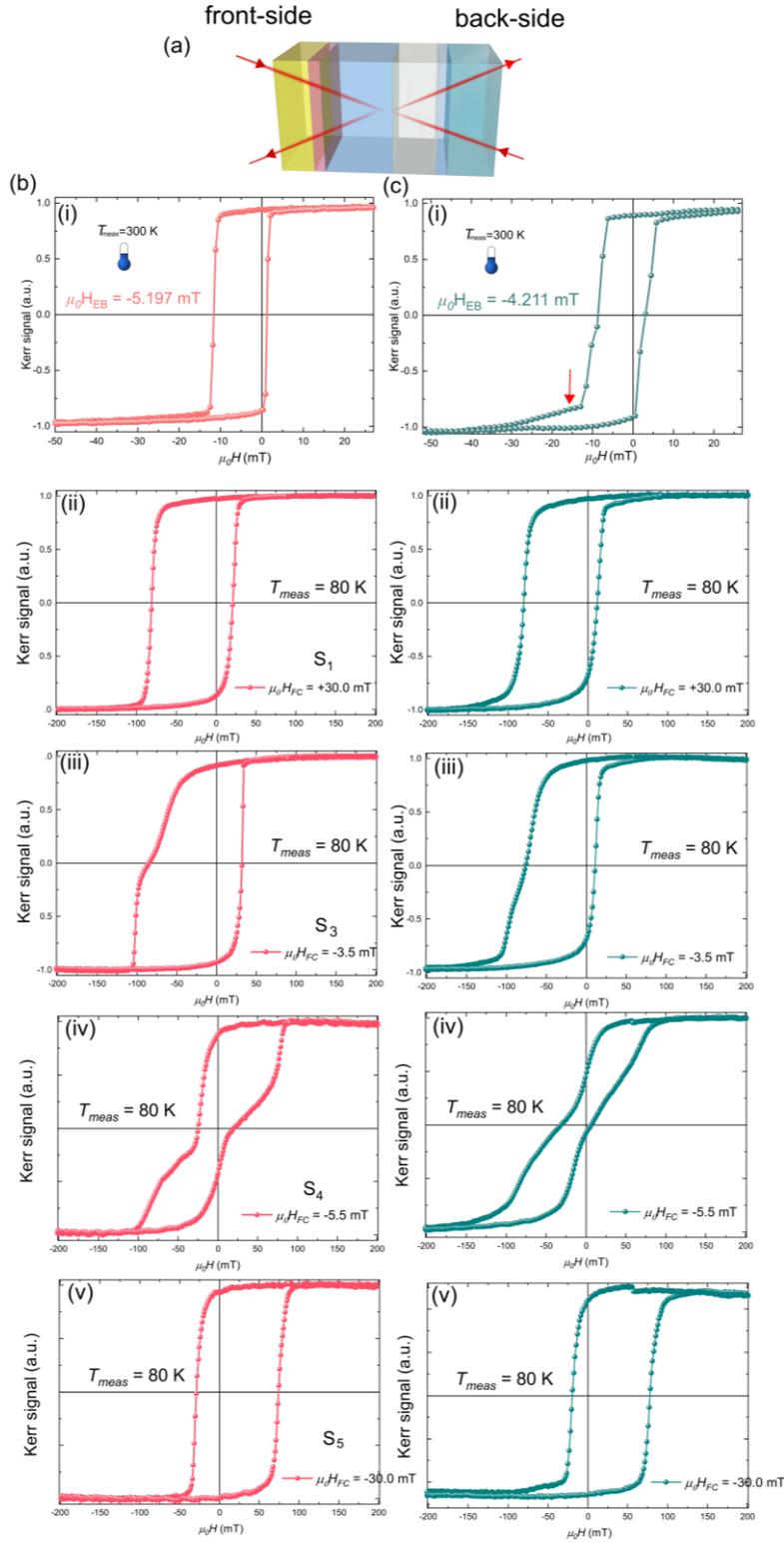

**Figure S5. Magneto-optical detection of MFC steady states. (a)** MOKE measurement geometry. For optical access from the front and back sides of the sample, a double-side polished MgO(001) substrate was used in the experiment. **(b)** The MOKE hysteresis loops obtained from the front side of the sample. **(b)(i)** The MOKE hysteresis loop obtained at 300 K indicates a negative EB field of the sample shown in Figure 6(a) of the main

paper. Due to the limited penetration depth of the MOKE laser, the MOKE measurements from the front side are not sensitive to the bottom $Ni_3Fe$(2.5 nm) switching process. **(b)(ii-v)** The MOKE hysteresis loops obtained at 80 K after field cooling the sample in different cooling states, as indicated in the plots. Very similar to the SQUID-VSM hysteresis loops shown in Figure 7b(i-v) in the main paper, a systematic transition of the magnetic hysteresis loop from initially negative ((b)(ii)) to a positive ((b)(v)) magnetic field direction can be seen, as a function of the applied cooling field ($\mu_0 H_{FC}$) sign and magnitude. **(c)(i)** The MOKE hysteresis loop obtained from the back side of the sample. In addition to the negative shift in the magnetic hysteresis loop (EB field), the contribution from the bottom NiFe(2.5 nm) layer can be seen, as indicated by a red arrow. **(c)(ii-v)** The MOKE hysteresis from the back-side of the sample obtained at 80 K in various cooling states, as indicated in the insets. The MOKE hysteresis loops shown in figures (b) and (c) confirms multiple-steady states of the sample in various cooling states. **Note**: All the MOKE hysteresis loops shown above were obtained from an average of 150 magnetic field cycles; therefore, we do not expect a contribution from the irreproducible EB training effect. We used a MOKE laser with a spot size of <10 µm, in comparison to SQUID-VSM measurements, where we used a 4 mm x 4 mm sample and measured the integral properties. Additionally, the sweep rate of the electromagnetic field can also have an influence on the measured $H_C$ and $H_{eb}$. Therefore, $H_C$ and $H_{eb}$ obtained using SQUID-VSM, and MOKE hysteresis loops differ. Nevertheless, as the same sample were used in both MOKE and SQUID-VSM measurements, the characteristic and systematic transition of the EB hysteresis loop shift from the initially set negative to a positive bias field direction is reproduced for various MFC states. The MFC procedure followed here to obtain the MOKE hysteresis loops in various cooling states is identical to the steps discussed in Figure 1(a) in the main paper.

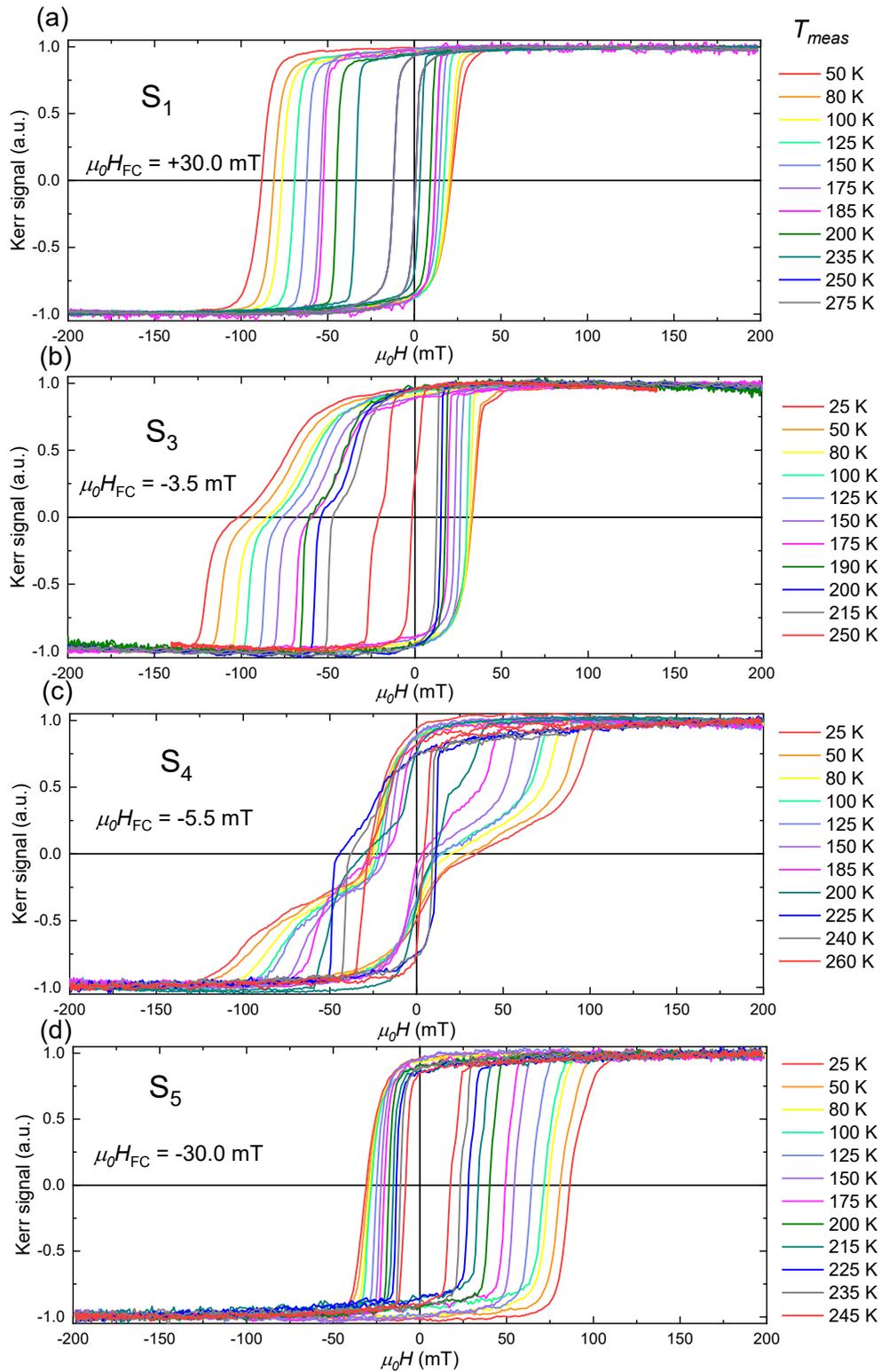

**Figure S6. Temperature-dependent MOKE hysteresis loops obtained in various MFC states from the front side of the sample. (a)** S$_1$ ($\mu_0 H_{FC}$ = +30 mT), **(b)** S$_2$ ($\mu_0 H_{FC}$ = -3.5 mT), **(c)** S$_4$ ($\mu_0 H_{FC}$ = -5.5 mT), and **(d)** S$_5$ ($\mu_0 H_{FC}$ = -30 mT) for the MgO(001)/Ni$_3$Fe(2.5 nm)/IrMn$_3$(5.2 nm)/Ni$_3$Fe(15 nm)/CoO(2.0 nm)/Pt(3.5 nm) sample. Due to the presence of an additional antiferromagnetic CoO layer, we observe a significant increase in

the coercive field and the EB field compared to the samples without the ultra-thin CoO layer, particularly below $T_N$ = 293 K. In states $S_1$ ($\mu_0 H_{FC}$ = +30 mT) and $S_5$ ($\mu_0 H_{FC}$ = +30 mT) from the magnetic hysteresis loops at 80 K; we obtain $\mu_0 H_{eb}$ = -29.878 mT and 23.228 mT, respectively. **Note:** The magnetic field cooling procedure followed here is identical to the steps shown and discussed in Figure 1 in the main paper.

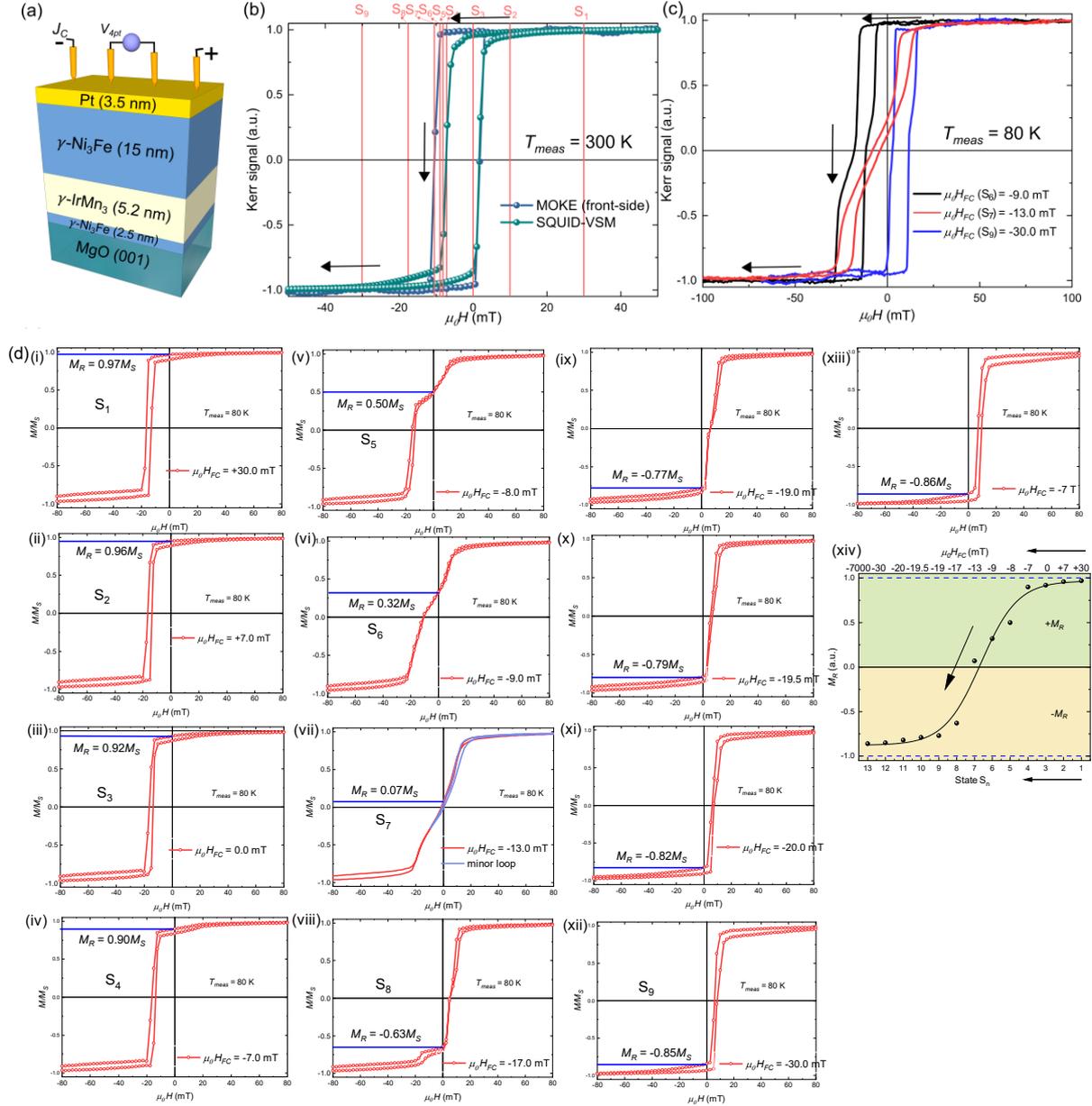

**Figure S7. Magnetic and magnetoresistance properties of MgO(001)/Ni₃Fe(2.5 nm)/IrMn₃(5.2 nm)/Ni₃Fe(15 nm)/Pt(3.5 nm) heterostructures in various MFC states. (a)** The sample structure and the magnetotransport measurement geometry ($J_C$∥$\mu_0H$∥[100]). Magneto-optical, SQUID-VSM, and magnetotransport measurements were performed on the same sample for a high degree of consistency. **(b)** Magnetic hysteresis loops obtained using the longitudinal MOKE magnetometry and SQUID-VSM at 300 K after field annealing and MFC the sample from 605 K to 300 K in +7 T magnetic field, followed by an EB training procedure. **(c)** The MOKE hysteresis loop obtained at 80 K after MFC the sample from 300 K to 80 K in various MFC states from the front side of the sample. The MOKE hysteresis loops were obtained from an average of 150 field cycles. **(d)(i)-(d)(xiii)** The SQUID-VSM hysteresis loops obtained at 80 K in various MFC states after field cycling (EB training) between

+7 T to -7 T and vice versa, as indicated. **(d)(xiv)** Change in the remanent magnetization $M_R$ in various MFC states obtained from the SQUID-VSM hysteresis loops. The EB field as a function of the applied cooling field is presented in Figure 9(a) of the main paper.

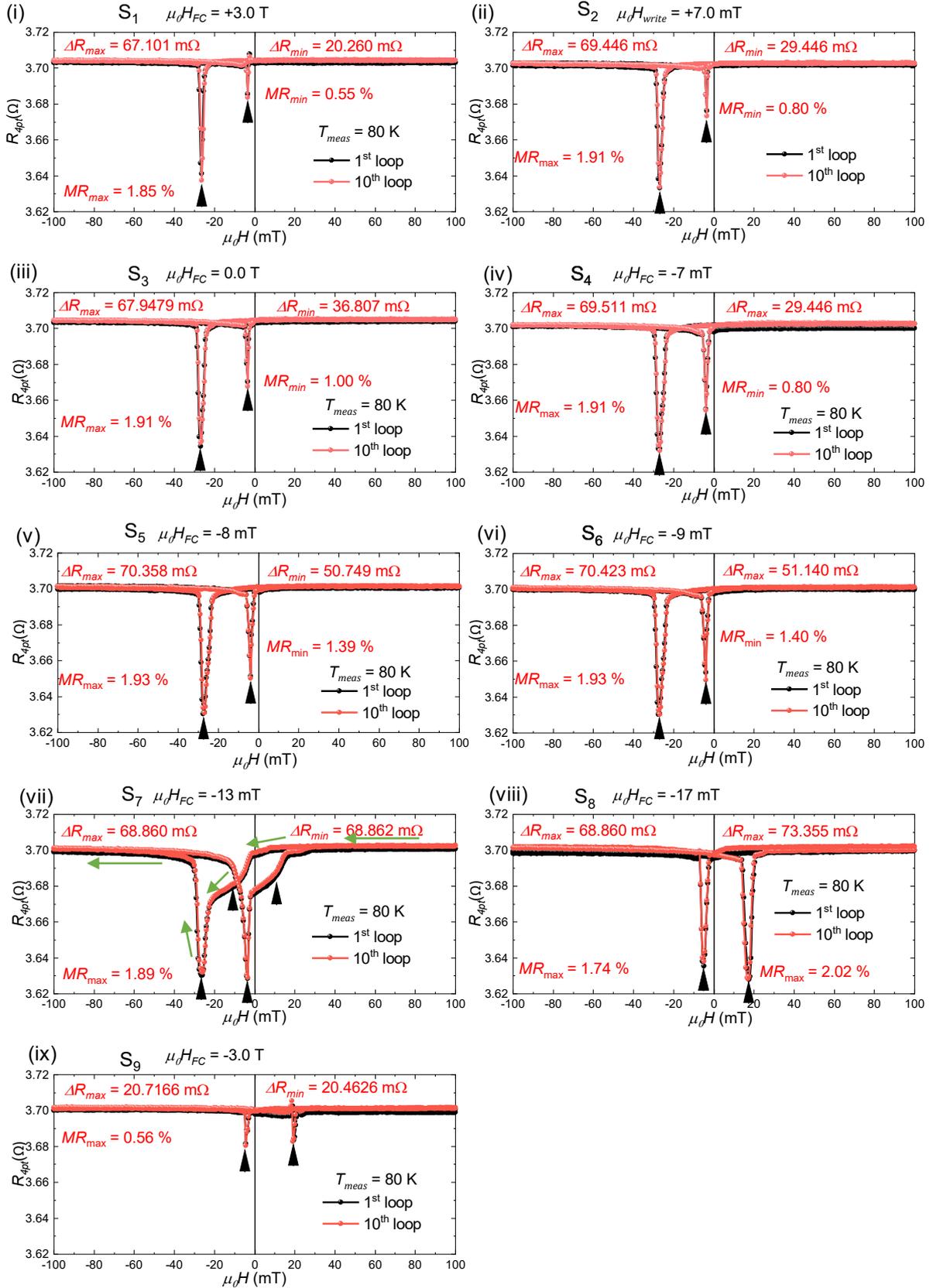

**Figure S7. (e)(i)-(d)(xi)** Magnetoresistive curves obtained in various MFC states (from state S₁-S₉) at 80 K. Arrowheads are a guide to the eye, indicating the negative resistance dips. The combination of MOKE, SQUID-

VSM, and magnetotransport measurements in various MFC states indicates a systematic transition of $H_{eb}$ from initially set negative to a positive bias field direction as a function of the applied cooling field. These collective experiments on the exchange coupled heterostructured sample further confirm that the domain state of the ferromagnetic layer during the iterative MFC process (from 300 K to 80 K) can be used to manipulate the AFM domain structure and the resulting systematic change in sign and magnitude of the exchange bias field. Unlike the heterostructures shown in Figure 6(a) of the main paper, the ultra-thin AFM CoO is removed. As a result, we see a reduction of $H_C$ and $H_{eb}$. Nevertheless, the characteristic EB field dependence on the FM domain states during the iterative MFC process persists. The MR curve shown in (vii) for MFC state S7 reveals an intermediate transition step. The portion of the FM $Ni_3Fe$(15 nm) thin film pinned along the positive magnetic field direction results in additional resistance dips indicated. Figure S7(d)(vii) shows the SQUID-VSM EB hysteresis loop of MFC state $S_7$; the FM magnetization (almost 45-50%) pinned along the positive magnetic field direction is switched via minor loop. The additional minor MR dips (or features) observed in (vii) denote the annihilation of the magnetic domains pinned (or biased) in the positive magnetic field direction. Intriguingly, even in MFC states, $S_5$ and $S_6$, the SQUID-VSM EB loops reveal a small portion of the magnetization (<30%) pinned in the positive magnetic field direction. Nevertheless, the corresponding MR curves for the MFC states $S_5$ (v) and $S_6$ (vi) did not reveal additional minor MR features.

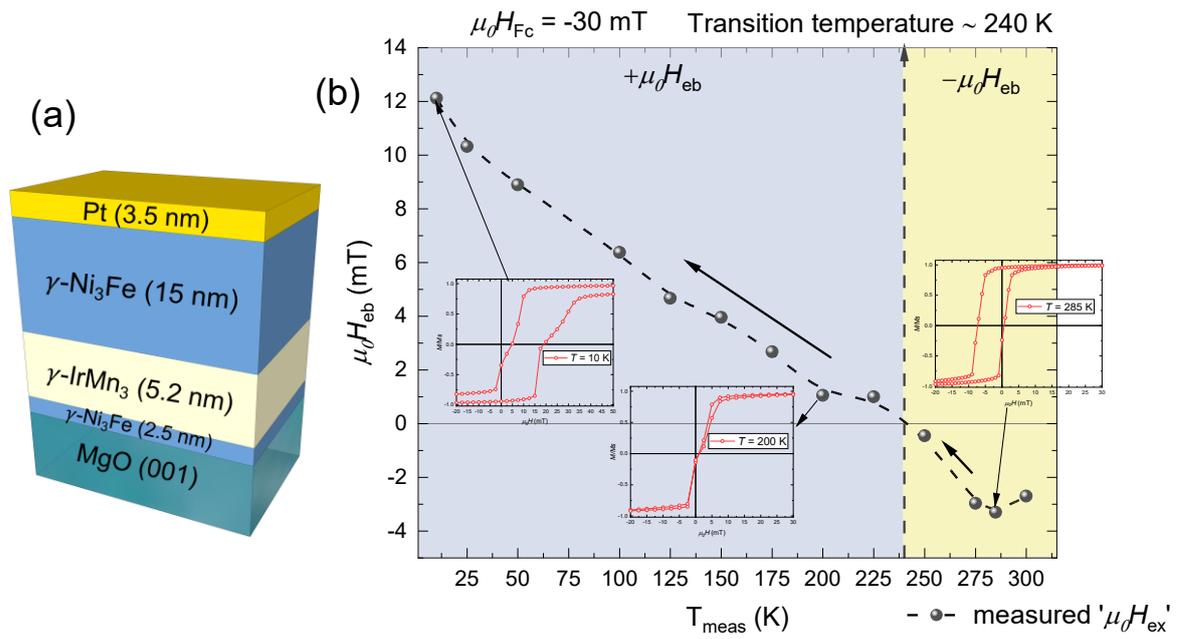

**Figure S8. Temperature-dependent transition of the EB field in MgO(001)/Ni₃Fe(2.5 nm)/IrMn₃(5.2 nm)/Ni₃Fe(15 nm)/Pt(3.5 nm) heterostructure.** (a) The layer structure of the sample. (b) $H_{eb}$ as a function of measurement temperature at a magnetic cooling field $\mu_0 H_{FC}$ = -30 mT. To determine the transition temperature where the sign of $H_{eb}$ changes from an initial set negative to a positive bias field direction, we applied $\mu_0 H_{FC}$ = -30 mT (opposite to the initially set bias field direction +7 T). We found that the sign change of $H_{eb}$ occurs approximately at 240 K. Note: The dashed lines are a guide to the eye.

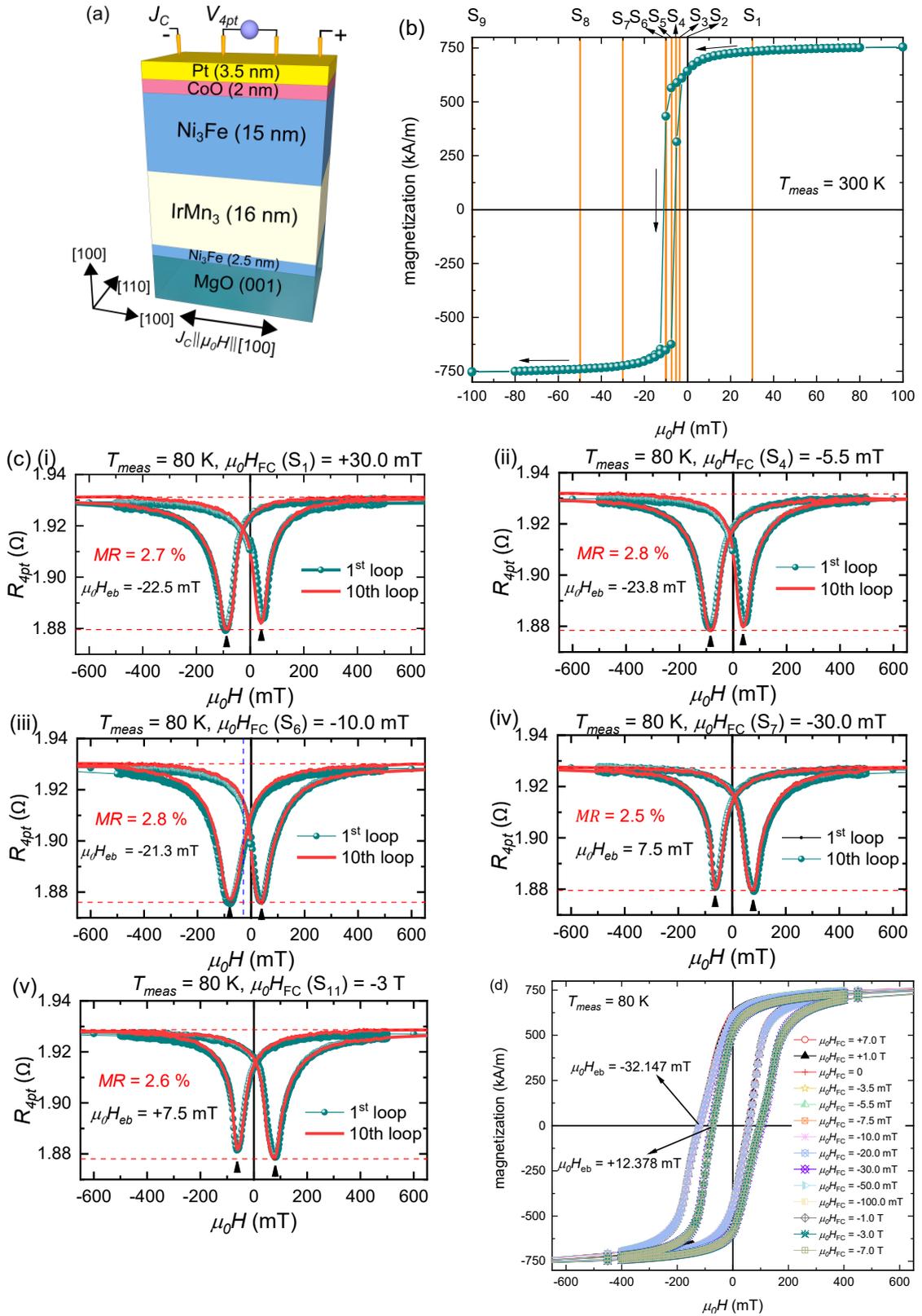

**Figure S9. Magnetic and magnetotransport properties of MgO(001)/Ni₃Fe(2.5 nm)/IrMn₃(16 nm)/Ni₃Fe(15 nm)/CoO(2.0 nm)/Pt(3.5 nm) heterostructures in various MFC states. (a)** The sample structure and the magnetotransport measurement geometry ($J_C$||$\mu_0 H$||[100]). (b) Magnetic hysteresis loops obtained using SQUID-VSM at 300 K after field annealing and MFC the sample from 605 K to 300 K in +7 T magnetic field, followed by an EB training procedure. **(c)(i)-(c)(iv)** Magnetoresistance curves obtained in various MFC states (from state $S_1$-$S_9$) at 80 K. Arrowheads are a guide to the eye, indicating the negative resistance dips. Compared to the heterostructures with IrMn(5.2 nm) shown in the main paper Figure (8), the change in resistance $\Delta R/R$ (MR) is smaller, and the difference between the individual MR states is relatively smaller, most likely due to an increase in the AFM grains with $V>V_C$, and larger, but fewer magnetic domains observed in films (see Supplementary Figure S9(e) below) **(d)** SQUID-VSM hysteresis loops obtained at 80 K in various MFC states as indicated. Note that the transition of the exchange bias field from initially negative (-32.14 mT) to positive (+12.37 mT) occurs sharply in MFC state $S_7$ (by overcoming the negative interfacial exchange energy), where $\mu_0 H_{FC}$ = -30 mT (magnetic field higher than the saturation field of the FM layer at 300 K). We achieved a certain degree of tunability in the system via iterative MFC due to the presence of disordered and reversible moments, mainly in the CoO(2.0 nm) layer. However, we expect that there are still a portion of grains in the IrMn₃(16 nm) thin film with $V<V_C$, prompted by the inevitable structure defects. In Supplementary Figure S10, we further study the MFC-dependent EB hysteresis properties of reference exchange coupled MgO(001)/Ni₃Fe (2.5 nm)/IrMn₃(16 nm)/Ni₃Fe(15 nm)/Pt(3.5 nm) heterostructures.

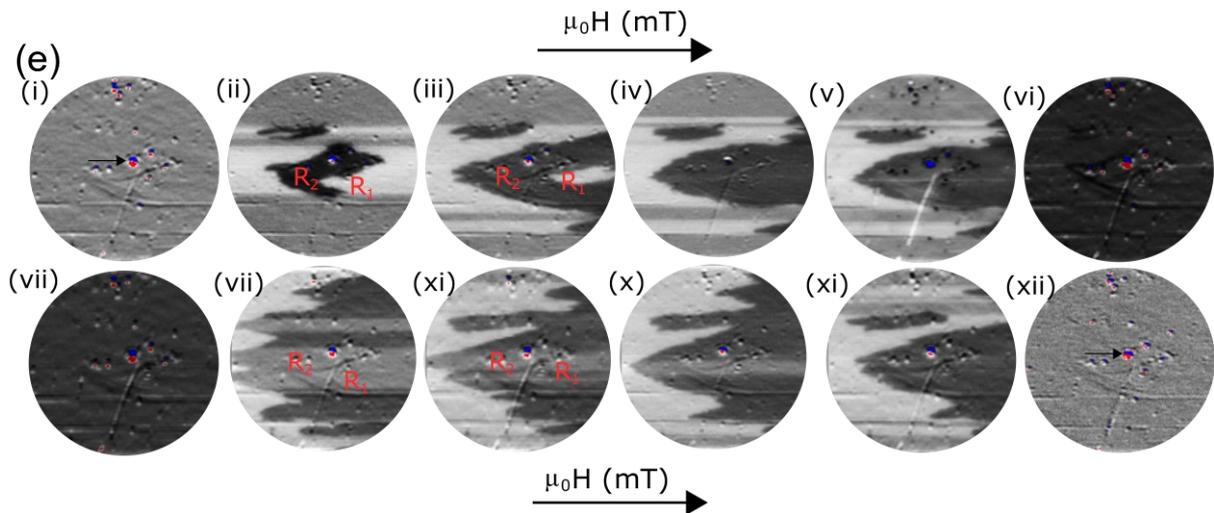

**Figure S9. (e) SKerrM image sequence for the sample shown in Figure S9(a) and the corresponding MOKE hysteresis loop is shown in Figure S9(b), the field of view is 1.2 mm in diameter. (i-vi) Magnetic domain structure obtained from the descending branch (EB direction) of the hysteresis loop shown in Figure S9(b). (vii-**

**xii)** Magnetic domain structure obtained from the ascending branch of the MOKE hysteresis loop shown in Figure S11(b). In contrast to the fractal domains observed in samples with $IrMn_3$(5.2 nm), we observe a large millimeter size domain structure and somewhat continuous domain wall motions due to the strain relaxation in the $Ni_3Fe$(15 nm) thin films. We found an asymmetric domain formation in ascending (EB direction) and descending branches of the MOKE hysteresis loop. The highlighted regions $\mathbf{R}_1$ and $\mathbf{R}_2$ in (ii & vii) and (iii & xi) show the asymmetric domains.

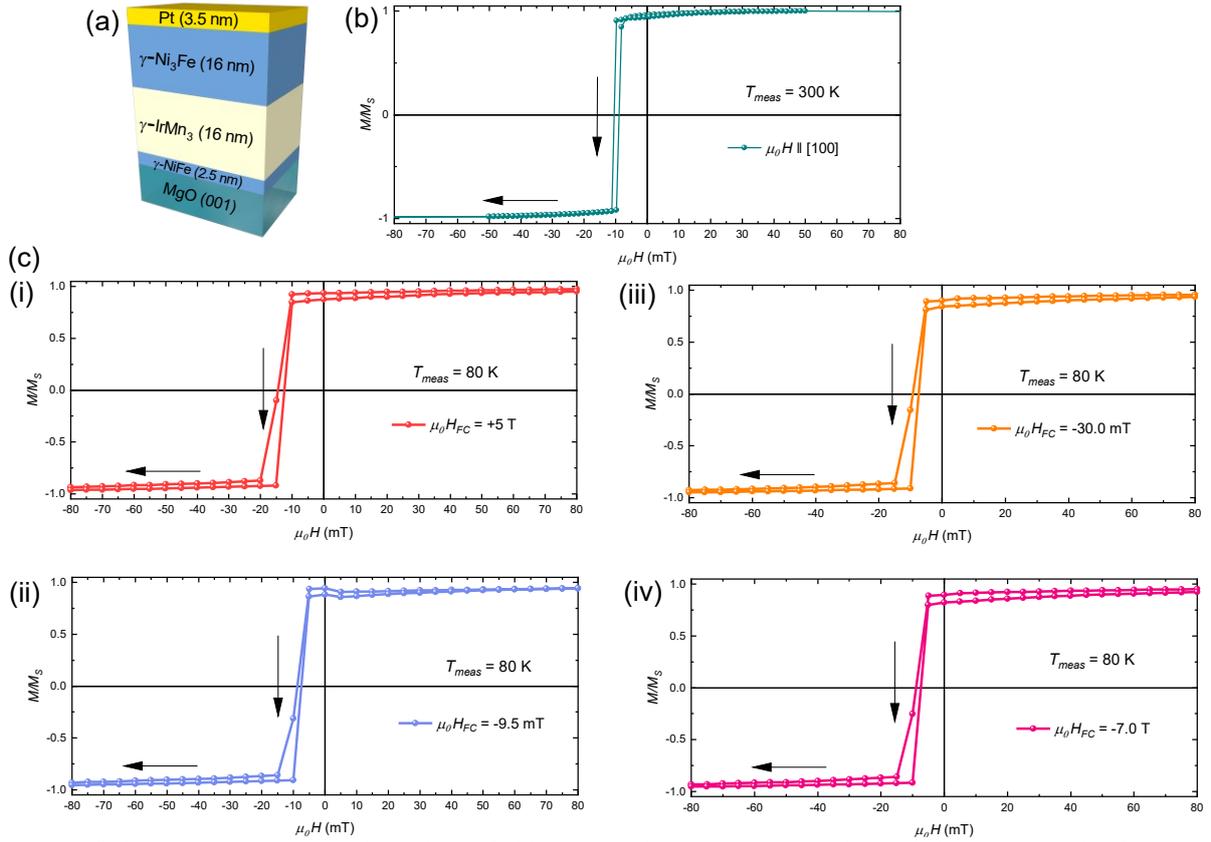

**Figure S10. Magnetic properties of MgO(001)/Ni₃Fe(2.5 nm)/IrMn₃(16 nm)/Ni₃Fe(15 nm)/Pt(3.5 nm) heterostructures in various MFC states. (a)** The sample structure. **(b)** Magnetic hysteresis loops obtained using SQUID-VSM at 300 K after field annealing and MFC the sample from 605 K to 300 K in +7 T magnetic field, followed by an EB training procedure. **(c)(i)-(c)(iv)** EB SQUID-VSM hysteresis loops obtained at 80 K in various MFC states as indicated. To check the stability of the EB field strength, initially, we obtained a hysteresis at 80 K after MFC the samples in +5 T (parallel to the initially set MFC direction +7 T). In (c)(i), we see a negative shift in the EB hysteresis loop. For instance, in (c)(iv), we field cooled the sample in -7 T (anti-parallel to the initial MFC direction), and the EB field remains in the negative field direction. However, we found a small reduction in the EB field strength compared to the EB loop shown in (c)(i). A summary of the EB field as a function of the applied cooling field is presented in the main paper, Figure 9(b). The reduction in the EB field strength when field cooled in the negative magnetic field is most likely caused by the persistence of the disordered and reversible IrMn₃ grains with $V<V_C$ even in films with IrMn₃(16 nm).

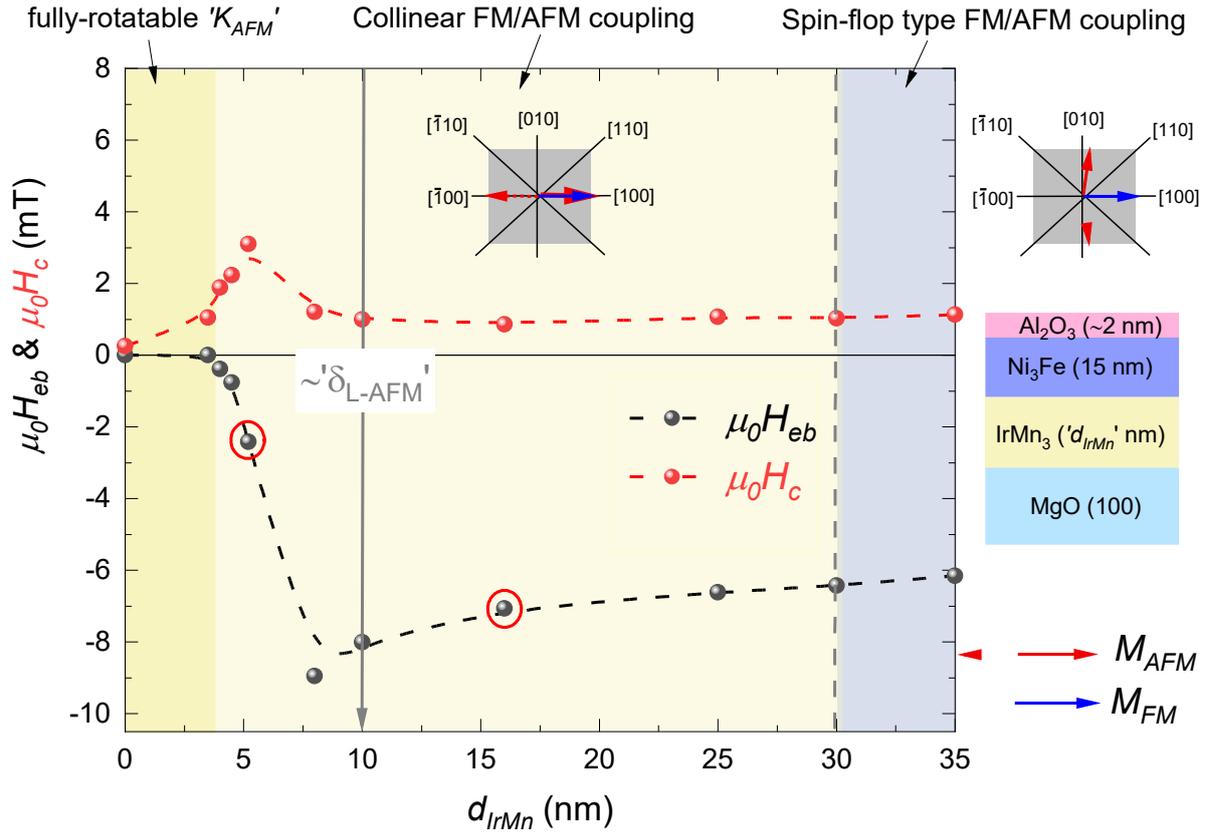

**Figure S11. Coercive and exchange bias field as a function of antiferromagnetic (ɤ-IrMn₃) film thickness in MgO(001)/IrMn₃/Ni₃Fe bilayers at 300 K.** $H_C$ and $H_{eb}$ as a function of AFM IrMn₃ thickness $d_{IrMn}$ at $T_{meas}$ = 300 K, extracted from the magnetic EB hysteresis loops of the bilayer samples shown in Supplementary Figure S12 below. At $d_{IrMn}$ = 4.0(±0.2) nm, we found the onset of the EB field. A critical AFM grain volume is needed to establish AFM order and induce the EB field at a given measurement temperature. At approximately $d_{IrMn}$ = 5.2 nm, we found a characteristic increase in the $H_C$, consistent with previous studies.[4,5] The enhancement in $H_C$ is found to be correlated with the presence of grains with complex AFM phases, namely disordered, reversible, frustrated, and irreversible order. We further found a maximum in the $H_{eb}$ at $d_{IrMn}$ = 8.0 nm, and beyond a critical AFM thickness, $H_C$ and $H_{eb}$ tend to saturate. All the samples with $d_{IrMn}$<30.0 nm showed a single-step reversal with a collinear alignment between the FM and the AFM interfacial spin structure. However, for samples with $d_{IrMn}$>30.0 nm, we found a spin-flop type interfacial exchange coupling between the FM and AFM spin structure.[6,7] A characteristic spin-flop EB hysteresis obtained at 300 K is shown in Supplement Figure S12(d).[8] **Note**: Dashed lines in the plot are a guide to the eye.

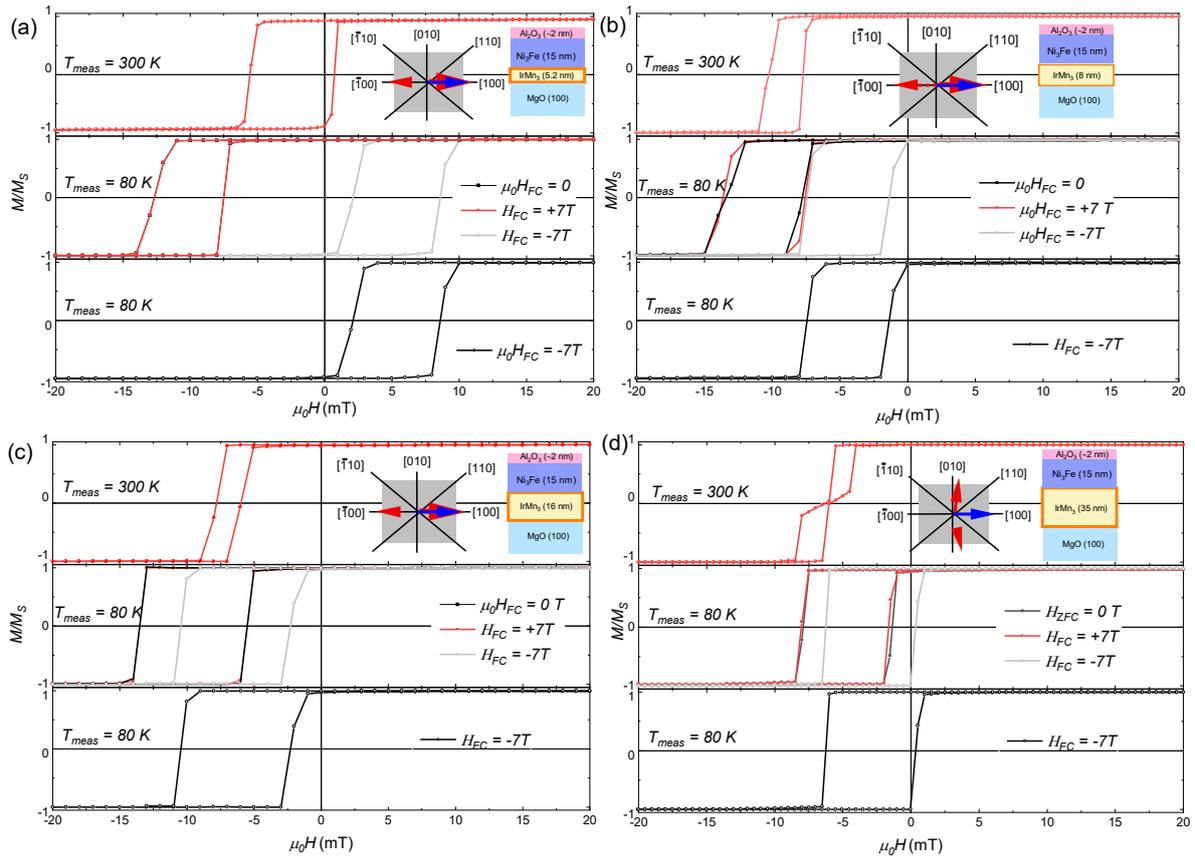

**Figure S12. Collinear and spin-flop type magnetization reversal behavior of MgO(001)/IrMn₃/Ni₃Fe bilayers as a function of antiferromagnetic (ɣ-IrMn₃) film thickness.** **(a)** EB hysteresis loop for AFM/FM bilayer sample with $d_{IrMn}$= 5.2 nm, obtained after MFC the samples in +7 T magnetic field at 300 K. To check the stability (and degeneracy) of the EB field, we MFC the samples down to 80 K in the various magnetic field configurations, either parallel or anti-parallel to the initial field annealing direction. We follow a similar approach for all the remaining AFM/FM bilayer samples **(b)** $d_{IrMn}$= 8.0 nm, **(c)** $d_{IrMn}$= 16 nm, and **(d)** $d_{IrMn}$= 35.0 nm. The sample with $d_{IrMn}$= 35.0 nm reveals a characteristic FM/AFM spin-flop EB hysteresis loop.[6-8] Regardless of the AFM thickness, all the samples showed a certain degree of degeneracy in the EB field strength when MFC from 300 K to 80 K in an anti-parallel magnetic field (-7 T) to the initial field annealed direction (+7 T). A summary of the $H_{eb}$ obtained at 80 K when MFC from 300 K to 80 K in parallel (+7 T) and anti-parallel (-7 T) is shown in Supplementary Figure S13 below. All the EB hysteresis loops were obtained after training the samples ten times in a magnetic field sequence (-7 T → +7 T→ -7 T→ +7 T).

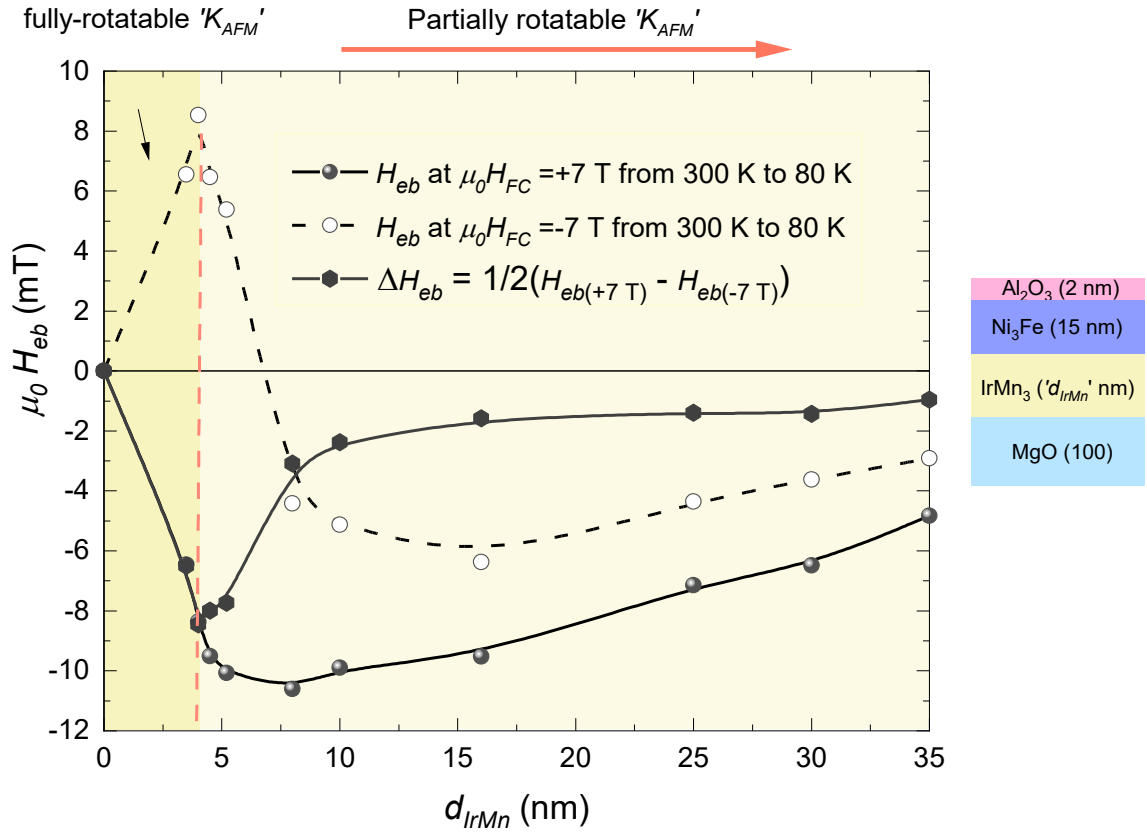

**Figure S13. Degeneracy of Exchange bias field as a function of antiferromagnetic (ɣ-IrMn₃) film thickness in MgO(001)/IrMn₃/Ni₃Fe bilayers at 80 K.** The degeneracy in the EB field strength when MFC from 300 K to 80 K in an anti-parallel magnetic field (-7 T) to the initial field annealed direction (+7 T). Independent of $d_{IrMn}$, we found that a certain degree of degeneracy persists. However, the magnitude of $\Delta H_{eb}$ strongly depends on the AFM IrMn grains ($V<V_C$) with disordered and reversible AFM-order. The iterative MFC procedure introduced in Figure 1(a) of the main paper can be tailored to control the orientation of the disordered and reversible AFM moments.

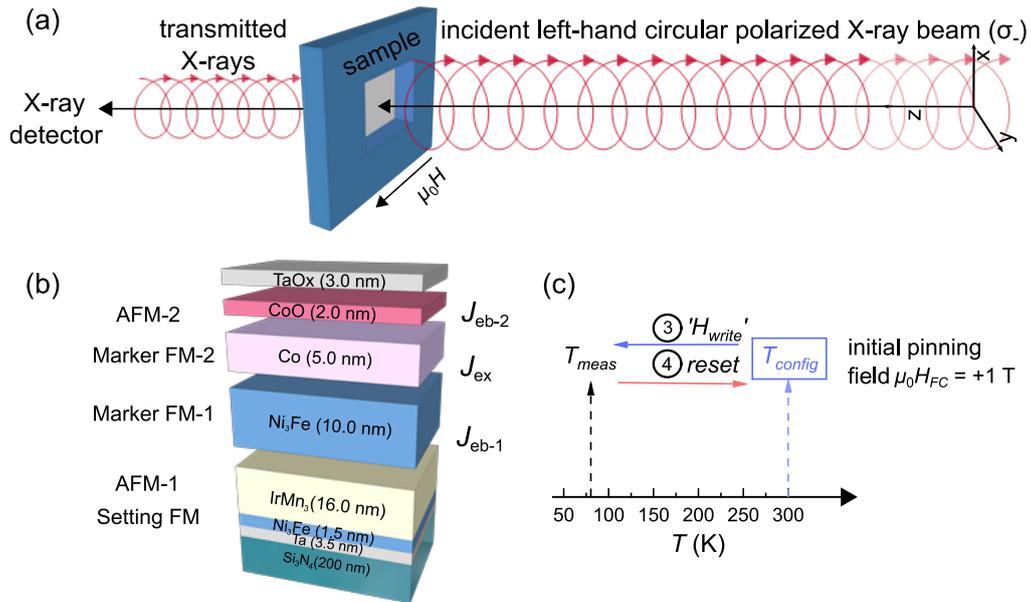

**Figure S14. Experimental geometry and the sample structure for XAS/XMCD experiments. (a)** The geometry of the measurement in transmission mode. The sample is tilted by 30° with respect to the incident x-ray beam-normal, essentially to probe the in-plane magnetic component. **(b)** Layer structure used in the element-specific XAS/XMCD measurements. **(c)** The iterative magnetic field cooling procedure used in the experiments. **Note**: In contrast to the samples investigated in the main paper, we introduced an additional FM Co(5.0 nm) layer at the top interface as a marker layer to probe the element-specific magnetization reversal behavior of the top FM/CoO interface. The total FM thickness remains at 15 nm, similar to the samples in the main paper. Moreover, we observed an FM exchange coupling between the Co/NiFe interfaces at 300 K.

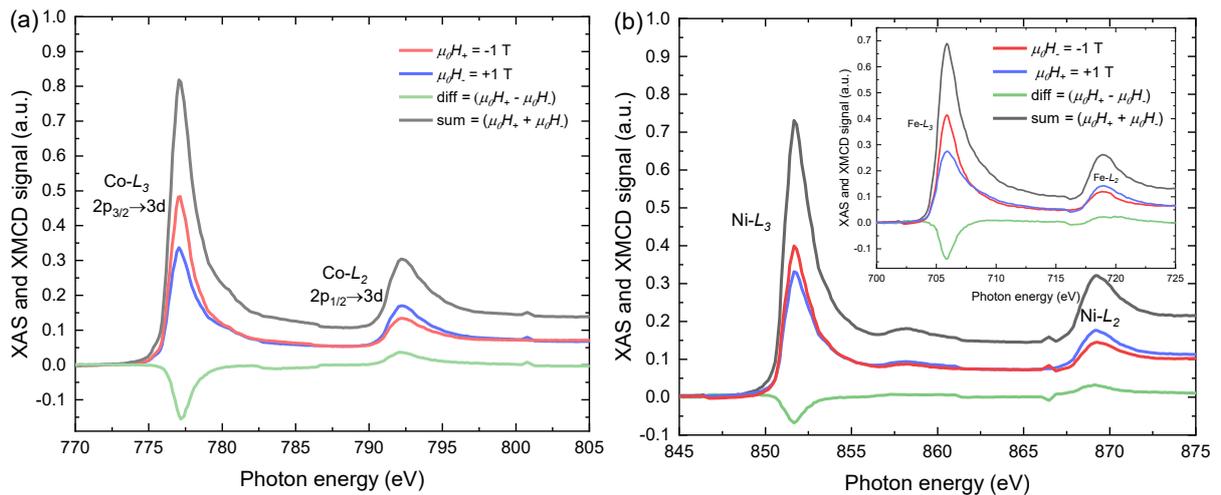

**Figure S15. XAS/XMCD spectra of Co, Ni, and Fe.** Selected energy absorption spectra of **(a)** Co and **(b)** Ni $L_{2,3}$-edges. The XMCD difference ($\mu_0H_+$ - $\mu_0H_-$) and the XAS sum ($\mu_0H_+$ + $\mu_0H_-$) signals are shown as well. The inset shows the XAS and XMCD signals for Fe $L_{2,3}$-edges. The spin polarization at the $L_3$ (spin-down) and $L_2$ (spin-up) edges have opposite signs due to the opposite spin-orbit coupling between the $2p_{3/2}$ and $2p_{1/2}$ levels.

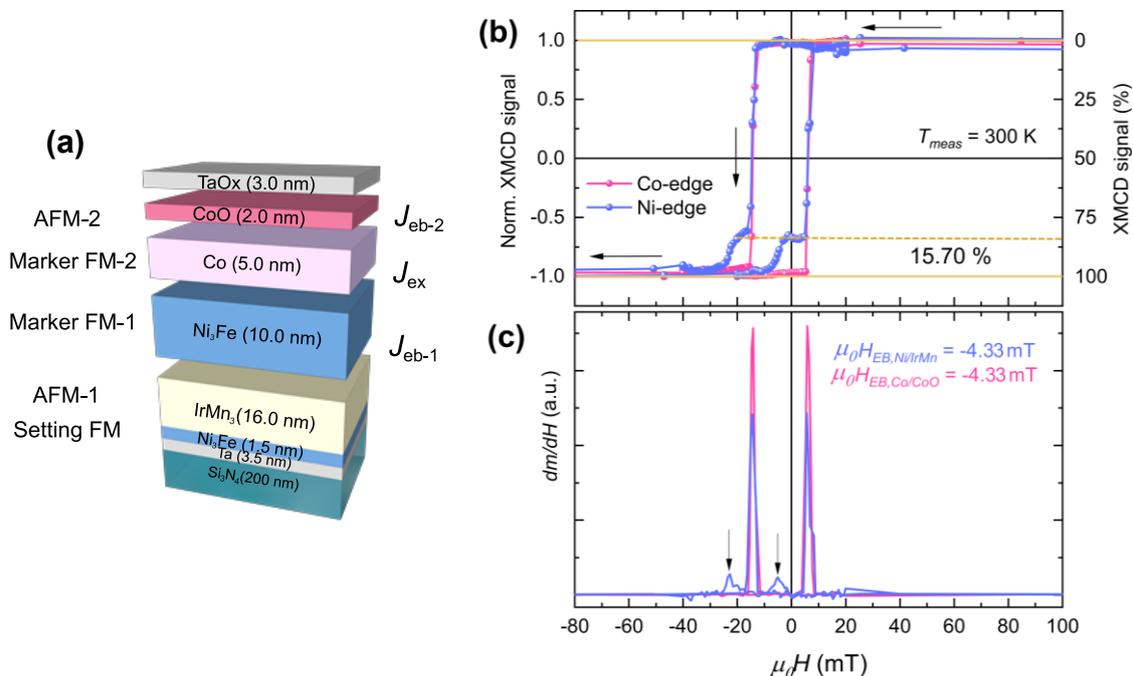

**Figure S16. Element-specific magnetic hysteresis loops obtained near the resonant Co and Ni $L_3$-edges at 300 K.**(a) Layer structure, and **(b and c)** Element-specific magnetic (XMCD) hysteresis loops and the corresponding first derivative measured at 300 K for the FM Ni ($E$ = 851.65 eV) and Co ($E$ = 776.91 eV) layers near the resonant $L_3$-edge. Note that the hysteresis loop at the Fe $L_3$-edge is skipped, as Fe magnetization resembles the characteristic behavior of Ni due to the strong ferromagnetic exchange interaction. The Ni hysteresis loop consists of contribution from the bottom and the top Ni/Mn interfaces, as confirmed by the two-step reversal behavior (absent in the Co hysteresis loop). Approximately 15.7 % of the XMCD signal accounts for the bottom seed Ni/Mn interface. We extract the EB field $\mu_0 H_{EB}$ = -4.33 mT from the loop shift of the Ni and Co magnetic hysteresis loops. At 300 K, the CoO remains in a paramagnetic state and does not contribute to the EB field.

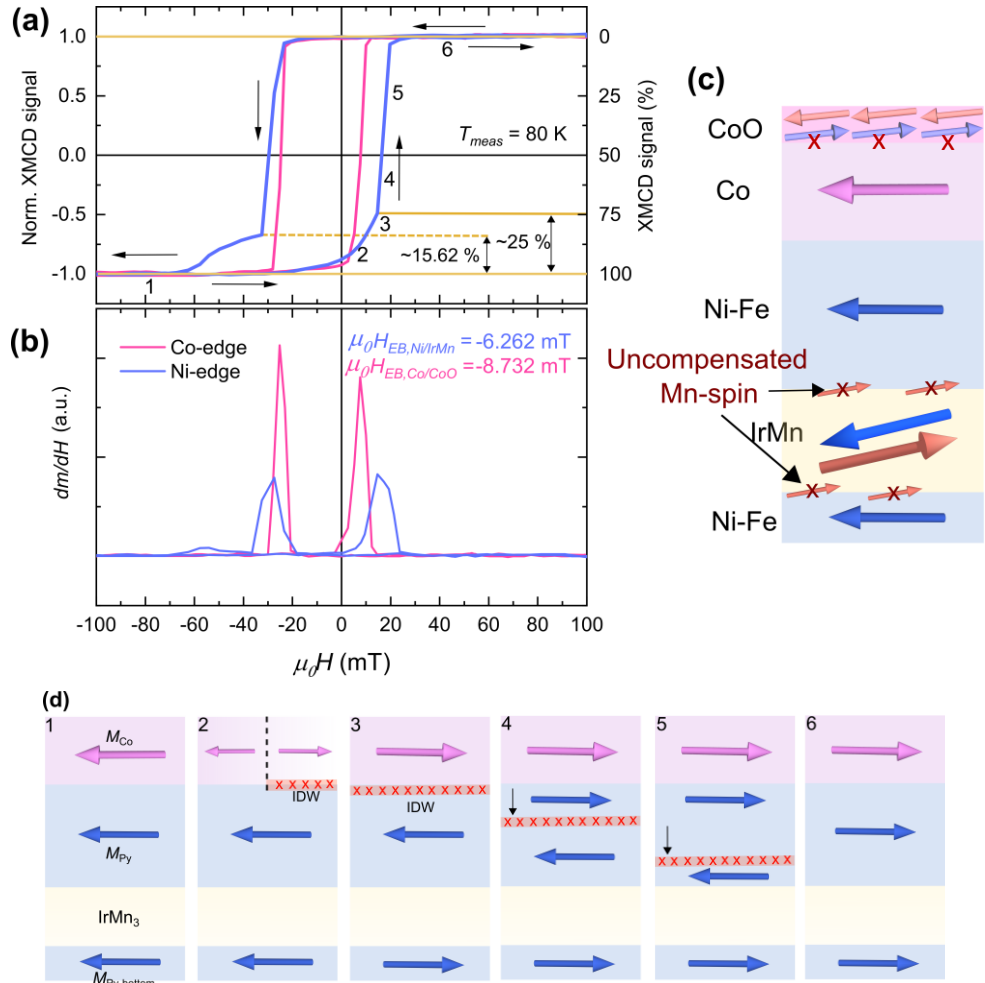

**Figure S17**. **Element-specific magnetic hysteresis loops obtained at 80 K in parallel MFC cooling state ($\mu_0H_{FC}$ = +100 mT).** **(a)** and **(b)** Element-specific magnetic (XMCD) hysteresis loops and the corresponding first derivative obtained at $T_{meas}$ = 80 K for Co ($E$ = 776.91 eV) and Ni ($E$ = 851.66 eV) $L_3$-edges. **(c)** Magnetic configuration of the sample at negative saturation marked in (a) as '**1**'. **(d)** Magnetic reversal (spin configuration of the Co and NiFe layers) of the sample in the ascending branch of the magnetic hysteresis loop (see (a) marked as 1-6). The magnetic reversal of the NiFe/Co occurs via nucleation and propagation of Néel-type interfacial domain wall (IDW). Such IDW permits the stabilization of anti-parallel alignment between the FM Co and NiFe layer (see positions marked as 3, 4, and 5 in (a and d)). As the applied cooling field is parallel to the initial pinning direction, we obtain a maximum EB field at Ni/Mn and Co/CoO interfaces. See Supplementary Table 1 for additional details. We obtain $\mu_0H_{eb}$=-6.262 mT at the top Ni/Mn interface and an interfacial exchange coupling

constant $J_{eb}$ = 47.76 μJ/m$^2$. The tail feature in the descending branch (biased direction) of the Ni hysteresis loop correspond to the bottom NiFe thin film contributions. At the Co/CoO interface we obtain $\mu_0 H_{eb}$ = -8.732 mT and $J_{eb1}$ = 59.90 μJ/m$^2$. We could not detect any measurable contribution (absorption signal) from the AFM CoO layer in the transmission geometry. Additionally, we capped the multilayers with a Ta+TaOx(3 nm) layer; therefore, the total electron yield detector has also proven ineffective in detecting the AFM CoO surface contribution.

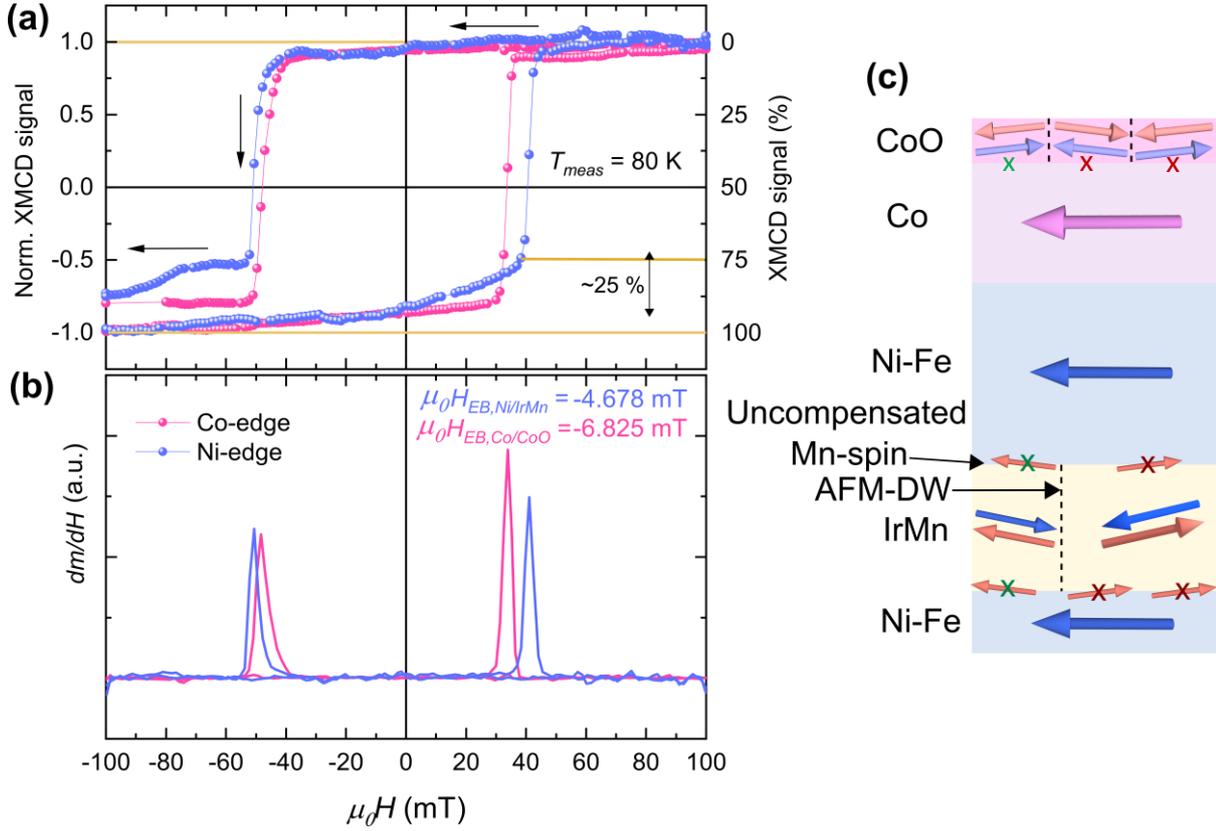

**Figure S18. Element-specific magnetic hysteresis loops obtained at 80 K in anti-parallel MFC cooling state ($\mu_0 H_{FC}$ = -100 mT). (a)** and **(b)** Element-specific magnetic (XMCD) hysteresis loops and the corresponding first derivative measured at $T_{meas}$ = 80 K for Co ($E$ = 777.91 eV) and Ni ($E$ = 851.67 eV) near the resonant $L_3$-edge. **(c)** The magnetic configuration of the sample at negative saturation. As indicated, the AFM domain walls are expected to be formed to separate the AFM grains with opposite spin orientations. Consequently, $|H_{eb}|$ at the Mn/Ni interface is ~25% less in the anti-parallel MFC state than the parallel MFC state. Moreover, we found an increase in $H_c$ at the Mn/Ni interface in the anti-parallel MFC state compared to the parallel MFC states shown in Figure S17. Similarly, at the top Co/CoO interface in the anti-parallel MFC state, we found a ~22% reduction in $|H_{eb}|$ and 153% increase in $H_C$ (see Supplementary Table 1 for additional details). The reduction in $H_{eb}$ at the top and bottom (AFM/FM: Mn/Ni and FM/AFM: Co/CoO) interface can be correlated to the significant modification of the AFM domain structure. As the thermally disordered and reversible AFM moments at the top and the bottom interface align anti-parallel with respect to the initially set (rigid) Mn-moments. Additionally, the dramatic increase in $H_C$ can be attributed to the stabilization of the domain walls perpendicular to the AFM/FM interfaces, as shown in (c) above.

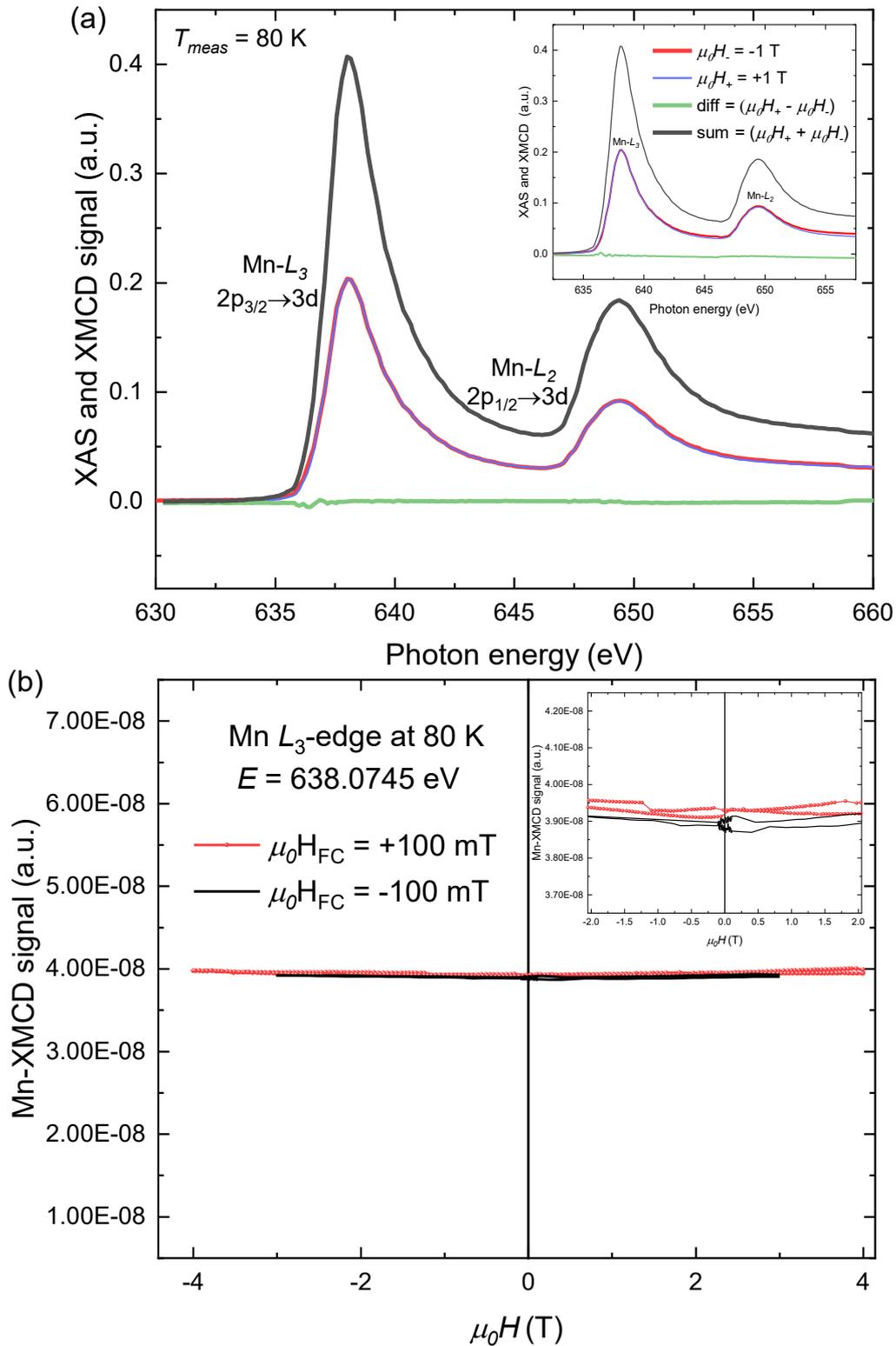

**Figure S19. Magnetic field stability of Antiferromagnetic Mn spin structure at 80 K in parallel and anti-parallel MFC states. (a)** Selected energy absorption spectra obtained at the Mn $L_{2,3}$-edges at 80 K after MFC the sample in $\mu_0H_{FC}$ = +100 mT (parallel MFC state). To check the magnetic field stability of the AFM Mn-moments after the MFC process at 80 K, we kept the photon helicity constant and obtained the absorption spectra at $\mu_0H_-$ =

-1 T and $\mu_0 H_+ = +1$ T, respectively. As the AFM Mn-moments are insensitive to the applied magnetic field, we did not see a detectable XMCD (difference: $\mu_0 H_+$-$\mu_0 H_-$) contrast. We heat the sample back to 300 K, and subsequently, field cooled the sample down to 80 K in -100 mT magnetic field (anti-parallel MFC state), opposite the initial field annealed direction. The inset in (a) shows the Mn $L_{2,3}$-edges XAS and XMCD spectra for $\mu_0 H_{FC} = -100$ mT. Similar to the previous case, we did not detect any visual XMCD contrast. The presence of Mauri-type AFM/FM planar domain walls at the Mn/Ni interface would cause a substantial change in the Mn absorption curve for the positive and negative applied magnetic field, as a portion of the AFM Mn-moments would rotate with the FM Ni and Fe moments.[9-11] However, the XMCD (difference) signal in either parallel or anti-parallel MFC states did not show such evidence. **(b)** Magnetic hysteresis loops (*as measured data, without normalization*) were obtained near the resonant AFM Mn $L_3$-edge ($E = 638.07$ eV) in parallel and anti-parallel (see inset) MFC states by applying a magnetic field of ±4 T. However, we did not observe any change in the in-plane Mn magnetization as a function of the applied magnetic field. Therefore, we infer that the Mn-moments are stable (or robust) against the applied magnetic field (within the measured magnetic field range ±4 T) after the MFC process (from 300 K to 80 K). Moreover, in the exchange coupled FM/AFM heterostructures studied here, Mauri-type interfacial (FM/AFM/FM/AFM) domain walls are highly unlikely to occur.[9,10] However, the domain wall perpendicular to the AFM/FM interface can still occur, as hinted by an increase in the coercive field at the Mn/Ni interface in the anti-parallel MFC state shown in Figure S18 and summary Table S1.

**Table S1.** Summary of various interfacial properties of the exchange coupled heterostructures measured at 300 K and 80 K, extracted from the element-specific XMCD hysteresis loops shown in Supplementary Figures S16 to S19.

| $T_{meas}$ (K) | Cooling state | $\mu_0 H_{FC}$ (T) | $\mu_0 H_C$ Ni/Mn (mT) | $\mu_0 H_{eb}$ Ni/Mn (mT) | $J_{eb1}$ Ni/Mn ($\mu$J/m$^2$) | $\mu_0 H_C$ Co/CoO (mT) | $\mu_0 H_{eb}$ Co/CoO (mT) | $J_{eb2}$ Co/CoO ($\mu$J/m$^2$) | $t_{IDW}$ (nm) |
|---|---|---|---|---|---|---|---|---|---|
| **300** | Initial state | 1 | 9.97[a] | -4.33[a] | 33.04[a] | 9.97 | n/a | n/a | n/a |
| | | | 8.81[b] | -14.02[b] | 14.93[b] | - | - | - | |
| **80** | Parallel | +0.1 | **22.56** | **-6.26** | 47.76 | 16.27 | -8.73 | 59.90 | ~1 |
| | Anti-parallel | -0.1 | **46.02** | **-4.67** | 35.67 | 41.15 | -6.82 | 46.82 | ~1 |

[a]parameters for the top Ni/Mn interface at 300 K, and [b]parameters for the bottom Ni/Mn interface. Note that the CoO remains in a paramagnetic state at 300 K and does not contribute to the exchange bias field. We calculated the effective exchange constant at Ni/IrMn and Co/CoO using the relation $J_{eb} = -\mu_0 H_{eb} M_{FM} d_{FM}$, where $M_{FM}$ and $d_{FM}$ are the saturation magnetization and nominal ferromagnet thickness, respectively.

# REFERENCES

(1) X. Zhang; Y. Lao; J. Sklenar; N. S. Bingham; J. T. Batley; J. D. Watts; C. Nisoli; C. Leighton; P. Schiffer. Understanding thermal annealing of artificial spin ice. APL Materials 7, 111112 (**2019**).

(2) T. Trunk; M. Redjdal; A. Kákay; M. F. Ruane; F. B. Humphrey. Domain wall structure in Permalloy films with decreasing thickness at the Bloch to Néel transition. J. Appl. Phys., 89, 7606 (**2001**).

(3) T. R. McGuire; R. I. Potter, Anisotropic magnetoresistance in ferromagnetic 3d alloys. IEEE Trans. Magn. 11, 1018–1038 (**1975**).

(4) M. Ali; C. H. Marrows; B. J. Hickey. Onset of exchange bias in ultrathin antiferromagnetic layers. Phys. Rev. B 67, 172405 (**2003**).

(5) M. Ali, C. H. Marrows, M. Al-Jawad, B. J. Hickey, A. Misra, U. Nowak, and K. D. Usadel. Antiferromagnetic layer thickness dependence of the IrMn/Co exchange-bias system. Phys. Rev. B 68, 214420 (**2003**).

(6) N. C. Koon. Calculations of Exchange Bias in Thin Films with Ferromagnetic/Antiferromagnetic Interfaces. Phys. Rev. Lett. 78, 4865 (**1997**).

(7) T. C. Schulthess; W. H. Butler. Consequences of Spin-Flop Coupling in Exchange Biased Films Phys. Rev. Lett. 81, 4516 (**1998**).

(8) W. Zhang; K. M. Krishnan. Spin-flop coupling and rearrangement of bulk antiferromagnetic spins in epitaxial exchange-biased Fe/MnPd/Fe/IrMn multilayers. Phys. Rev. B 86, 054415 (**2012**).

(9) D. Mauri; H. C. Siegmann; P. S. Bagus; E. Kay. Simple model for thin ferromagnetic films exchange coupled to an antiferromagnetic substrate. J. Appl. Phys. 62, 3047 (**1987**).

(10) Joo-Von Kim; R. L. Stamps. Hysteresis from antiferromagnet domain-wall processes in exchange-biased systems: Magnetic defects and thermal effects. Phys. Rev. B 71, 094405 (**2005**).

(11) P. Wadley; K. W. Edmonds; M. R. Shahedkhah; R. P. Campion; B. L. Gallagher; J. Železný; J. Kuneš; V. Novák; T. Jungwirth; V. Saidl; P. Němec; F. Maccherozzi; S. S. Dhesi. Control of antiferromagnetic spin axis orientation in bilayer Fe/CuMnAs films. Sci Rep 7, 11147 (**2017**).